\documentclass[11pt,a4paper]{article}
\pdfoutput=1

\usepackage{jheppub}
\usepackage{amsmath}
\usepackage{amssymb}
\usepackage{graphics}
\usepackage[active]{srcltx}
\usepackage{pdfsync}
\usepackage{shuffle}
\usepackage{slashed}
\usepackage{hyperref}
\usepackage{subfigure}

\newcommand{\bea}{\begin{eqnarray}}
\newcommand{\eea}{\end{eqnarray}}
\newcommand{\nn}{\nonumber \\}

\def\W #1{\widetilde{#1}}

\def\eref#1{(\ref{#1})}

\def\a{{\alpha}}

\def\b{{\beta}}

\allowdisplaybreaks

\title{Towards New Hidden Zero and $2$-Split of Loop-Level Feynman Integrands in ${\rm Tr}(\phi^3)$ Model}
\author[a]{Kang Zhou}
\affiliation[a]{Center for Gravitation and Cosmology, College of Physical Science and Technology, Yangzhou University,\\
No.180, Siwangting Road, Yangzhou, 225009, P.R. China}

\emailAdd{zhoukang@yzu.edu.cn}
\date{\today}
\abstract{
We extend the hidden zeros and $2$-split of tree-level ${\rm Tr}(\phi^3)$ amplitudes to loop-level Feynman integrands, apart from some physically irrelevant scaleless integrals. Our method is based on a certain factorization mechanism that occurs in Feynman diagrams when summing over shuffle permutations. The loop-level hidden zeros and $2$-split identified in this work differ from those in the literature. In our result, the kinematic conditions for loop-level hidden zeros and $2$-split are remarkably simple. Their connection is as tight as at tree-level, with the same procedure for obtaining the $2$-split condition from the zero condition. The resulting $2$-split formula at loop-level represents a generalization of that at tree-level: the $L$-loop integrand is expressed as a sum over $L+1$ terms, each of which exhibits a $2$-split structure.
}


\begin{document}

\maketitle \flushbottom

%%%%%%%%%%%%%%%%%%%%%%%%%%%%%%
\section{Introduction}
\label{sec-intro}
%%%%%%%%%%%%%%%%%%%%%%%%%%%%%%%%

In the past few years, the discovery of two novel properties at tree level---hidden zeros and $2$-split---has emerged as a remarkable advancement in the study of scattering amplitudes. Hidden zeros refer to the phenomenon that on special loci in kinematic space, the entire amplitude vanishes identically. First identified in the ${\rm Tr}(\phi^3)$, non-linear sigma model (NLSM) and Yang-Mills (YM) theories \cite{Arkani-Hamed:2023swr}, hidden zeros have since been shown to hold across a wider range of models, including the special Galileon, Dirac-Born-Infeld theory, gravity (GR), as well as modified YM and GR with specific higher-derivative corrections \cite{Rodina:2024yfc,Bartsch:2024amu,Li:2024qfp,Zhang:2024efe,Huang:2025blb,Zhou:2025tvq}. By slightly altering these zero loci, another striking property emerges: the $2$-split behavior, where the amplitude factorizes exactly into a product of two lower-point off-shell currents, without the need for taking residue at a pole \cite{Cao:2024gln,Cao:2024qpp,Arkani-Hamed:2024fyd}.
These two new properties were initially discovered using modern frameworks such as the associahedron, surfaceology and Cachazo-He-Yuan (CHY) formalism \cite{Arkani-Hamed:2023swr,Cao:2024gln,Cao:2024qpp,Arkani-Hamed:2024fyd}. In subsequent work, efforts have also been made to understand them from various different perspectives, such as universal expansions of amplitudes, BCFW recursion relation, and traditional Feynman diagrams \cite{Zhou:2024ddy,Feng:2025ofq,Feng:2025dci}. For further research on hidden zeros and $2$-split of tree amplitudes, see \cite{Guevara:2024nxd,Cao:2025hio,Zhang:2025zjx,Zhang:2026dcm,Saha:2026ftv,Azevedo:2025vxo,CarrilloGonzalez:2026lnu}.

The importance of these novel structures extends far beyond mere computational curiosities. They reveal that tree-level amplitudes obey constraints stronger than those dictated by locality and unitarity alone, pointing toward an unexplored geometric foundation of the S-matrix. Moreover, the universality of hidden zeros and $2$-split across scalars, gauge bosons, and gravitons suggests a common kinematic origin, providing fresh insights into the unification of different models. From a practical standpoint, the $2$-split property offers a new computational tool: it reduces higher-point amplitudes to products of lower-point currents, thereby simplifying calculations and enabling new bootstrap strategies. Furthermore, these two newly discovered properties provide novel avenues for addressing the boundary term problem in BCFW recursion relation and for deriving soft theorems of tree amplitudes \cite{Li:2025suo,Zhou:2025xly,Zhou:2026isc}.

Given these remarkable tree-level features, a natural and pressing question is whether these two properties persist at loop-level.
In \cite{Backus:2025hpn}, hidden zeros of $1$-loop ${\rm Tr}(\phi^3)$ and NLSM Feynman integrands were studied by means of the kinematic mesh at $1$-loop-level. On the other hand, in \cite{Arkani-Hamed:2024fyd}, the $2$-split of the loop-level Feynman integrands for ${\rm Tr}(\phi^3)$ was investigated using surfaceology, and its generalizations to NLSM and YM theories were discussed.

In this paper, using the Feynman-diagram-based method developed in our previous work \cite{Zhou:2024ddy}, we find new hidden zeros and $2$-split of loop-level ${\rm Tr}(\phi^3)$ Feynman integrands that differ from those given in the literature \cite{Backus:2025hpn,Arkani-Hamed:2024fyd}. Our approach is based on a factorization mechanism exhibited by Feynman diagrams, which we refer to as shuffle factorization along a specific line (SFASL). More concretely, under the kinematic constraints of the hidden zeros or the $2$-split, when various building blocks of Feynman diagrams are attached to a specific line, the result of summing over shuffle permutations factorizes into two independent parts\footnote{A similar shuffle factorization also appears in recent studies of cosmological wavefunctions \cite{Li:2026gns}.}. A more detailed explanation of this mechanism will be given in the following sections. In \cite{Zhou:2024ddy}, this mechanism was used to interpret hidden zeros and $2$-split of tree ${\rm Tr}(\phi^3)$ amplitudes. This mechanism is completely local, concerning only the interaction vertices on a specific line and independent of the structures of other parts of the Feynman diagrams. This feature allows it to apply equally well at loop-level. Therefore, in this paper, we utilize this mechanism to generalize hidden zeros and $2$-split to loop-level.

Compared with the hidden zeros and $2$-split of loop-level integrands found in the literature, the kinematic conditions under which the hidden zeros and $2$-split are realized in our new results are simpler and weaker. Moreover, the connection between the kinematic condition for hidden zeros and that for $2$-split remains as tight as at tree-level. In particular, the procedure for obtaining the kinematic condition for $2$-split by relaxing the condition for zeros is exactly the same as that at tree-level. The loop-level $2$-split we obtain is a generalization of its tree-level counterpart. More specifically, after imposing the kinematic condition for $2$-split given in this work, the entire $L$-loop Feynman integrand reduces to a sum of $L+1$ terms, each of which exhibits the $2$-split structure.

The remainder of this paper is organized as follows. In section \ref{sec-tree}, we review the interpretation of hidden zeros and $2$-split developed in our previous work \cite{Zhou:2024ddy}, from the Feynman diagram perspective. In particular, we provide a detailed introduction to the mechanism exhibited by Feynman diagrams, which we refer to as SFASL. Based on the diagrammatic method introduced in section \ref{sec-tree}, we extend the hidden zeros and $2$-split to the $1$-loop Feynman integrand of ${\rm Tr}(\phi^3)$ model in section \ref{sec-1loop}. Next, in section \ref{sec-higherloop}, we establish the hidden zeros and $2$-split at higher loops using the same approach. Finally, we present a brief conclusion and discussion in section \ref{sec-conclu}.

%%%%%%%%%%%%%%%%%%%%%%%%%%%%%%%
\section{Review for hidden zero and $2$-split of tree-level amplitudes in ${\rm Tr}(\phi^3)$ model}
\label{sec-tree}
%%%%%%%%%%%%%%%%%%%%%%%%%%%%%%%

In this section, we review the interpretation for hidden zeros and $2$-split of tree ${\rm Tr}(\phi^3)$ amplitudes from the Feynman diagram perspective. This material does not arise from the work presented in this paper, but rather from our previous work \cite{Zhou:2024ddy}. Nevertheless, since this Feynman-diagram-based approach serves as the foundation for studying hidden zeros and $2$-split of loop-level Feynman integrands in subsequent sections, it is necessary to introduce it in detail. The organization and emphasis of this section is significantly different from that of \cite{Zhou:2024ddy}: we will highlight a mechanism exhibited by Feynman diagrams in the summation over shuffle permutations, which we refer to as SFASL. This mechanism will play a key role in the study of loop-level integrands in the following sections.

Before proceeding to the review, we first give a brief introduction to color-ordered ${\rm Tr}(\phi^3)$ amplitudes at tree-level. The ${\rm Tr}(\phi^3)$ model describes the cubic interaction of colored massless scalars. Its Lagrangian is given by
\bea
{\cal L}_{{\rm Tr}(\phi^3)}={\rm Tr}(\partial\phi)^2+g\,{\rm Tr}(\phi^3)\,,
\eea
where $\phi$ is an $N\times N$ matrix, carrying one index in the fundamental representation of
$SU(N)$ and another in the anti-fundamental representation. The color-ordered tree-level amplitudes in this model, with coupling constants stripped off, consist solely of massless scalar propagators.

In subsection \ref{subsec-tree-example}, we will provide a detailed introduction to SFASL and present two simple examples. In subsection \ref{subsec-tree-proofshuffle}, we will give a general proof of SFASL using a recursive approach. In subsection \ref{subsec-tree-zerosplit}, we will interpret hidden zeros and $2$-split of tree ${\rm Tr}(\phi^3)$ amplitudes through this mechanism.

%%%%%%%%%%%%%%%%%%%%%%%%%%%%%%%
\subsection{Shuffle permutation and factorization along specific line: two simple examples}
\label{subsec-tree-example}
%%%%%%%%%%%%%%%%%%%%%%%%%%%%%%%

%
\begin{figure}
  \centering
   \includegraphics[width=8cm]{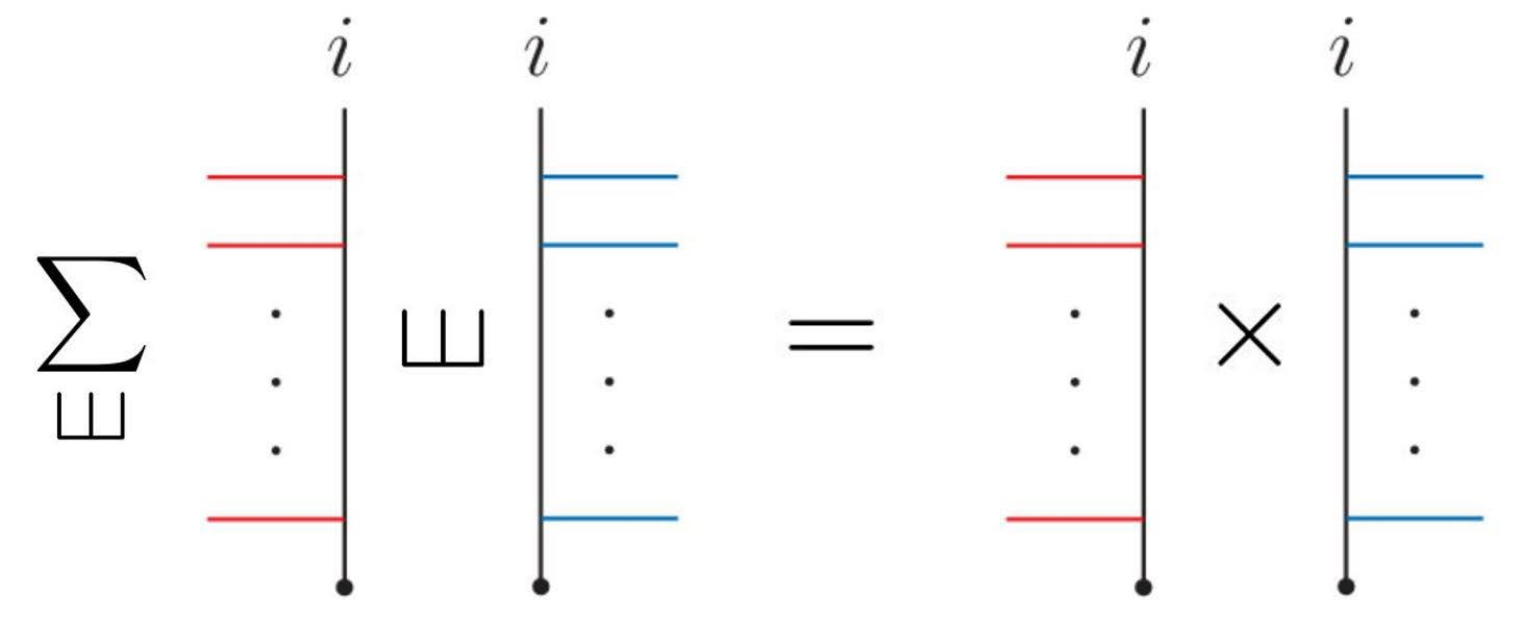} \\
  \caption{Factorization along the line $L_{i,\bullet}$ happens when summing over shuffle permutations. The red lines are $A$-lines, while the blue lines are $B$-lines.}\label{Fig1}
\end{figure}

\begin{figure}
  \centering
   \includegraphics[width=12cm]{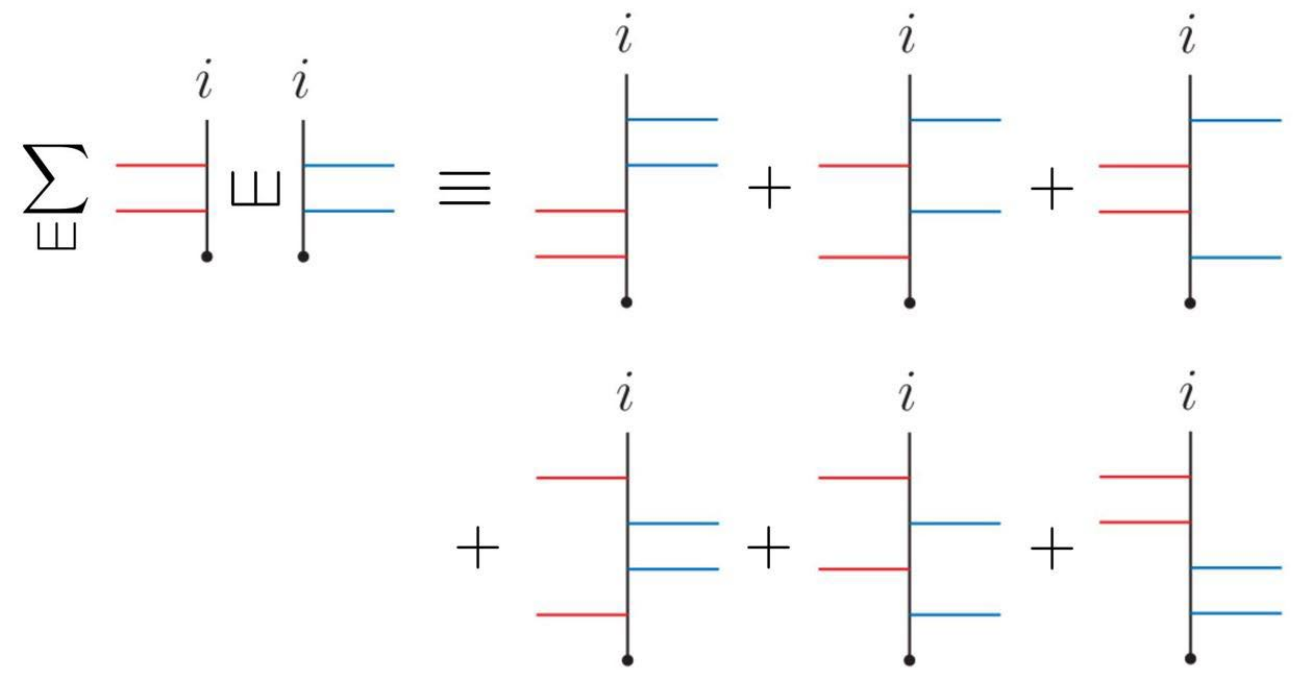} \\
  \caption{The meaning of summing over shuffle permutations.}\label{Fig2}
\end{figure}

When certain kinematic constraints are satisfied, Feynman diagrams exhibit a non-trivial behavior during the summation over different diagrams, which we refer to as shuffle factorization along a specific line (SFASL). Such behavior is illustrated in Fig.\ref{Fig1}. We refer to each red line in this figure as an $A$-line, and each blue line as a $B$-line. For an $A$-line labeled $a$ and a $B$-line labeled $b$, their momenta satisfy
\bea
k_a\cdot k_b=0\,.~~~~\label{kine-condi-shuffle}
\eea
Each $A$-line or $B$-line is attached to a special line $L_{(i,\bullet)}$ via a cubic vertex of the ${\rm Tr}(\phi^3)$ model. In the line $L_{(i,\bullet)}$, $i$ is a massless external leg satisfying the on-shell condition $k_i^2=0$, while $\bullet$ is a cubic vertex.
As usual, the summation over shuffles $\shuffle$ means summing over all permutations subject to preserving the relative order of the $A$-lines and the relative order of the $B$-lines. An instance of such summation is shown in Fig.\ref{Fig2}. Note that although the diagrams on both sides of $\shuffle$ each contain the line $L_{(i,\bullet)}$, in the actual meaning of summing over shuffle permutations, there is only one such line $L_{(i,\bullet)}$, as illustrated in Fig.\ref{Fig2}. Based on the above interpretation, Fig.\ref{Fig1} shows that when the $A$-lines and $B$-lines satisfy the kinematic constraint \eref{kine-condi-shuffle}, the result of summing over all shuffles factorizes into the product of the two components participating in the shuffle. This is what we call factorization along the special line $L_{(i,\bullet)}$. We emphasize that each $A$-line or $B$-line can be either an external line or an internal line, and the momentum it carries can be either on-shell or off-shell.

\begin{figure}
  \centering
   \includegraphics[width=10cm]{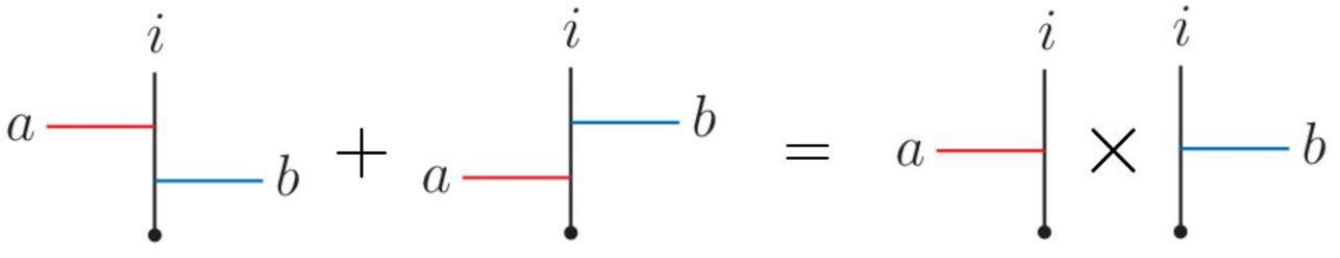} \\
  \caption{The first example of Fig.\ref{Fig1}.}\label{Fig3}
\end{figure}

To illustrate more explicitly the SFASL, let us consider two concrete examples.
The first example is shown in Fig.\ref{Fig3}. In this graph, the l.h.s. is the summation over the shuffles of $a$ and $b$, namely,
\bea
{1\over s_{ia}}\,{1\over s_{iab}}+{1\over s_{ib}}\,{1\over s_{iab}}\,,~~~\label{lhs-fig3}
\eea
where $s_{ia}=(k_i+k_a)^2$, $s_{ib}=(k_i+k_b)^2$, and $s_{iab}=(k_i+k_a+k_b)^2$. Using the kinematic condition \eref{kine-condi-shuffle}, as well as the on-shell condition $k_i^2=0$, we ge the following simple but useful observation:
\bea
s_{iab}\,\xrightarrow[]{\eref{kine-condi-shuffle}}\,\big(2\,k_i\cdot k_a+k_a^2\big)+\big(2\,k_i\cdot k_b+k_b^2\big)=s_{ia}+s_{ib}\,.~~~\label{key-observation}
\eea
Plugging this observation into \eref{lhs-fig3}, we immediately find
\bea
{1\over s_{ia}}\,{1\over s_{iab}}+{1\over s_{ib}}\,{1\over s_{iab}}={1\over s_{ia}}\,{1\over s_{ib}}\,.~~\label{meaning-fig3}
\eea
This is precisely what Fig.\ref{Fig3} means.

\begin{figure}
  \centering
   \includegraphics[width=14cm]{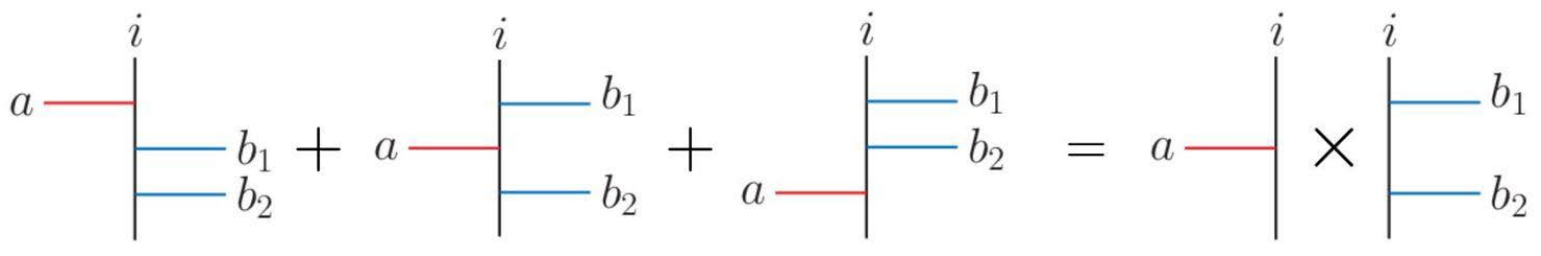} \\
  \caption{The second example of Fig.\ref{Fig1}.}\label{Fig4}
\end{figure}

The second example is given in Fig.\ref{Fig4}. The l.h.s. of this figure is the summation over shuffles of $a$ and $\{b_1,b_2\}$, which can be expressed as
\bea
{1\over s_{ia}}\,{1\over s_{iab_1}}\,{1\over s_{iab_1b_2}}+{1\over s_{ib_1}}\,{1\over s_{iab_1}}\,{1\over s_{iab_1b_2}}
+{1\over s_{ib_1}}\,{1\over s_{ib_1b_2}}\,{1\over s_{iab_1b_2}}\,.~~~~\label{lhs-fig4}
\eea
Applying the previous conclusion \eref{meaning-fig3} to the first and second terms in \eref{lhs-fig4}, we get
\bea
{1\over s_{ia}}\,{1\over s_{iab_1}}\,{1\over s_{iab_1b_2}}+{1\over s_{ib_1}}\,{1\over s_{iab_1}}\,{1\over s_{iab_1b_2}}
={1\over s_{ia}}\,{1\over s_{ib_1}}\,{1\over s_{iab_1b_2}}\,.
\eea
Combining this with the third term in \eref{lhs-fig4}, we obtain the following factorization behavior
\bea
{1\over s_{ib_1}}\,{1\over s_{ia}}\,{1\over s_{iab_1b_2}}+{1\over s_{ib_1}}\,{1\over s_{ib_1b_2}}\,{1\over s_{iab_1b_2}}
={1\over s_{ia}}\,\Big({1\over s_{ib_1}}\,{1\over s_{ib_1b_2}}\Big)\,,
\eea
where we have used
\bea
s_{iab_1b_2}\,\xrightarrow[]{\eref{kine-condi-shuffle}}\,\big(2\,k_i\cdot k_a+k_a^2\big)+\Big(2\,k_i\cdot (k_{b_1}+k_{b_2})+(k_{b_1}+k_{b_2})^2\Big)=s_{ia}+s_{ib_1b_2}\,,
\eea
which is the direct generalization of the observation \eref{key-observation}.
We have therefore established the following factorization behavior,
\bea
{1\over s_{ia}}\,{1\over s_{iab_1}}\,{1\over s_{iab_1b_2}}+{1\over s_{ib_1}}\,{1\over s_{iab_1}}\,{1\over s_{iab_1b_2}}
+{1\over s_{ib_1}}\,{1\over s_{ib_1b_2}}\,{1\over s_{iab_1b_2}}={1\over s_{ia}}\,\Big({1\over s_{ib_1}}\,{1\over s_{ib_1b_2}}\Big)\,,
~~~\label{meaning-fig4}
\eea
which is precisely the meaning of Fig.\ref{Fig4}.

Recalling the derivation of \eref{meaning-fig4}, we began with the earlier result \eref{meaning-fig3} and proved recursively the factorization structure shown in Fig.\ref{Fig4}. This recursive pattern offers inspiration for the general proof of the factorization behavior illustrated in Fig.\ref{Fig1}. In the next subsection, we shall pursue this idea to prove the SFASL.

%%%%%%%%%%%%%%%%%%%%%%%%%%%%%%%
\subsection{Shuffle permutation and factorization along specific line: recursive proof}
\label{subsec-tree-proofshuffle}
%%%%%%%%%%%%%%%%%%%%%%%%%%%%%%%

In this subsection, we provide a recursive proof for the general SFASL illustrated in Fig.\ref{Fig1}.

To present the proof more clearly, we first introduce some notation.
When the numbers of $A$-lines and $B$-lines are $p$ and $q$, respectively, we denote the summation over shuffles along the line $L_{(i,\bullet)}$ as
as
\bea
\sum_{\shuffle(p,q)}\,\prod_{t=1}^{p+q}\,{1\over D_t^{(i,\bullet)}}\,,
\eea
where $1/D_t^{(i,\bullet)}$ are propagators along $L_{(i,\bullet)}$. As two examples,
summations \eref{lhs-fig3} and \eref{lhs-fig4} can be understood as
\bea
\sum_{\shuffle(1,1)}\,\prod_{t=1}^{1+1}\,{1\over D_t^{(i,\bullet)}}&=&{1\over s_{ia}}\,{1\over s_{iab}}+{1\over s_{ib}}\,{1\over s_{iab}}\,,\nn
\sum_{\shuffle(1,2)}\,\prod_{t=1}^{1+2}\,{1\over D_t^{(i,\bullet)}}&=&{1\over s_{ia}}\,{1\over s_{iab_1}}\,{1\over s_{iab_1b_2}}+{1\over s_{ib_1}}\,{1\over s_{iab_1}}\,{1\over s_{iab_1b_2}}
+{1\over s_{ib_1}}\,{1\over s_{ib_1b_2}}\,{1\over s_{iab_1b_2}}\,.
\eea
Using this notation, the SFASL illustrated in Fig.\ref{Fig1} can be expressed as
\bea
\sum_{\shuffle(p,q)}\,\prod_{t=1}^{p+q}\,{1\over D_t^{(i,\bullet)}}\,=\,\Big(\prod_{\a=1}^p\,{1\over s_{ia_1\cdots a_{\a}}}\Big)\,
\Big(\prod_{\b=1}^q\,{1\over s_{ib_1\cdots b_{\b}}}\Big)\,,~~\label{fac-propa-gen}
\eea
where $A$-lines and $B$-lines are encoded as $\{a_1,\cdots, a_p\}$ and $\{b_1,\cdots,b_q\}$, respectively. When $p=0$ or $q=0$, we also define
\bea
\sum_{\shuffle(p,0)}\,\prod_{t=1}^{p}\,{1\over D_t^{(i,\bullet)}}\,&=&\,\prod_{\a=1}^p\,{1\over s_{ia_1\cdots a_{\a}}}\,,\nn
\sum_{\shuffle(0,q)}\,\prod_{t=1}^{q}\,{1\over D_t^{(i,\bullet)}}\,&=&\,\prod_{\b=1}^q\,{1\over s_{ib_1\cdots b_{\b}}}\,.~~\label{p=0-q=0}
\eea
The graphical meaning of \eref{p=0-q=0} is manifest since no shuffle is required when $p=0$ or $q=0$.

In the previous subsection, we have proved the factorization behavior \eref{fac-propa-gen} for $(p,q)=(1,1)$. Meanwhile, when $p=0$ or $q=0$, the relation in \eref{p=0-q=0} automatically holds. These results serve as the starting point for the recursion. To finish the proof, let us assume that the SFASL in \eref{fac-propa-gen} is valid for $(p,q-1)$ and $(p-1,q)$. Then for $(p,q)$ we have
\bea
\sum_{\shuffle(p,q)}\,\prod_{t=1}^{p+q}\,{1\over D_t^{(i,\bullet)}}&=&\Big[\Big(\sum_{\shuffle(p,q-1)}\,\prod_{t=1}^{p+q-1}\,{1\over D_t^{(i,\bullet)}}\,\Big)\,+\Big(\sum_{\shuffle(p-1,q)}\,\prod_{t=1}^{p+q-1}\,{1\over D_t^{(i,\bullet)}}\,\Big)\,\Big]\,{1\over s_{ia_1\cdots a_{p}b_1\cdots b_{q}}}\nn
&=&\Big[\Big(\prod_{\a=1}^{p}\,{1\over s_{ia_1\cdots a_{\a}}}\Big)\,
\Big(\prod_{\b=1}^{q-1}\,{1\over s_{ib_1\cdots b_{\b}}}\Big)\,+\Big(\prod_{\a=1}^{p-1}\,{1\over s_{ia_1\cdots a_{\a}}}\Big)\,
\Big(\prod_{\b=1}^{q}\,{1\over s_{ib_1\cdots b_{\b}}}\Big)\,\Big]\,{1\over s_{ia_1\cdots a_{p}b_1\cdots b_{q}}}\nn
&=&\Big(\prod_{\a=1}^{p-1}\,{1\over s_{ia_1\cdots a_{\a}}}\Big)\,
\Big(\prod_{\b=1}^{q-1}\,{1\over s_{ib_1\cdots b_{\b}}}\Big)\,\Big[{1\over s_{1a_1\cdots a_{p}}}+{1\over s_{1b_1\cdots b_{q}}}\Big]\,{1\over s_{ia_1\cdots a_{p}b_1\cdots b_{q}}}\nn
&=&\Big(\prod_{\a=1}^{p-1}\,{1\over s_{ia_1\cdots a_{\a}}}\Big)\,\Big(\prod_{\b=1}^{q-1}\,{1\over s_{ib_1\cdots b_{\b}}}\Big)\,{1\over s_{1a_1\cdots a_{p}}}\,{1\over s_{1b_1\cdots b_{q}}}\nn
&=&\Big(\prod_{\a=1}^{p}\,{1\over s_{ia_1\cdots a_{\a}}}\Big)\,\Big(\prod_{\b=1}^{q}\,{1\over s_{ib_1\cdots b_{\b}}}\Big)\,.~~\label{proof-fac-propa}
\eea
In the above, the second step uses the assumption that \eref{fac-propa-gen} holds for $(p,q-1)$ and $(p-1,q)$, while the fourth step uses the following generalization of observation \eref{key-observation},
\bea
s_{ia_1\cdots a_{p}b_1\cdots b_{q}}\,&\xrightarrow[]{\eref{kine-condi-shuffle}}&\,\Big[2\,k_i\cdot \Big(\sum_{\a=1}^{p}\,k_{a_\a}\Big)+\Big(\sum_{\a=1}^{p}\,k_{a_\a}\Big)^2\Big]+\Big[2\,k_i\cdot \Big(\sum_{\b=1}^{q}\,k_{b_\b}\Big)+\Big(\sum_{\b=1}^{q}\,k_{b_\b}\Big)^2\Big]\nn
&=&s_{ia_1\cdots a_{p}}+s_{ib_1\cdots b_{q}}\,,
\eea
which implies
\bea
{1\over s_{1a_1\cdots a_{p}}}+{1\over s_{1b_1\cdots b_{q}}}={s_{ia_1\cdots a_{p}b_1\cdots b_{q}}\over s_{ia_1\cdots a_{p}}\,\,s_{ib_1\cdots b_{q}}}\,.
\eea

The result in \eref{proof-fac-propa} demonstrates that if the SFASL \eref{fac-propa-gen} holds for $(p, q-1)$ and $(p-1, q)$, then it necessarily holds for $(p, q)$. The recursive proof is thus completed.

Evidently, the proof presented in this subsection depends only on interactions localized on the line $L_{(i,\bullet)}$. It means, the factorization along the line $L_{i,\bullet}$, depicted in \eref{fac-propa-gen} and Fig.\ref{Fig1}, does not depend at all on whether the $A$-lines or $B$-lines are internal or external, nor on what components the $A$-lines or $B$-lines are attached to. In the next subsection, we will show that when the $A$-lines and $B$-lines are connected to tree-level currents, the SFASL \eref{fac-propa-gen} leads to hidden zeros and $2$-split of tree amplitudes. In subsequent sections, we will see that when the $A$-lines or $B$-lines contain internal lines in the loops, the SFASL \eref{fac-propa-gen} yields hidden zeros and $2$-split of loop-level Feynman integrands.

%%%%%%%%%%%%%%%%%%%%%%%%%%%%%%%
\subsection{Hidden zero and $2$-split from shuffle factorization along specific line}
\label{subsec-tree-zerosplit}
%%%%%%%%%%%%%%%%%%%%%%%%%%%%%%%

We now explain that the SFASL depicted in \eref{fac-propa-gen} and Fig,\ref{Fig1} directly leads to hidden zeros of tree-level ${\rm Tr}(\phi^3)$ amplitudes and the $2$-splits near the zeros.

Let us begin with hidden zeros. For an $n$-point tree ${\rm Tr}(\phi^3)$ amplitude, one can choose a pair of external legs $i$ and $j$, and denote legs between them as two ordered sets $\pmb A=\{i+1,\cdots,j-1\}$ and $\pmb B=\{j+1,\cdots,i-1\}$. The hidden zero for the certain choice of $(i,j)$
states that
\bea
{\cal A}_n(1,\cdots,n)\,\to&\,0\,,~~~~~~{\rm if}~k_{\hat{a}}\cdot k_{\hat{b}}=0\,,
~~~~{\rm for}~\forall\,\,\hat{a}\in\pmb A\,,~\hat{b}\in\pmb B\,.~~\label{zero-tree}
\eea
In the above, we labeled on-shell legs as $\hat{a}$ and $\hat{b}$, to distinguish them from lines $a$ and $b$ in \eref{kine-condi-shuffle} which can be either on-shell or off-shell.

\begin{figure}
  \centering
   \includegraphics[width=12cm]{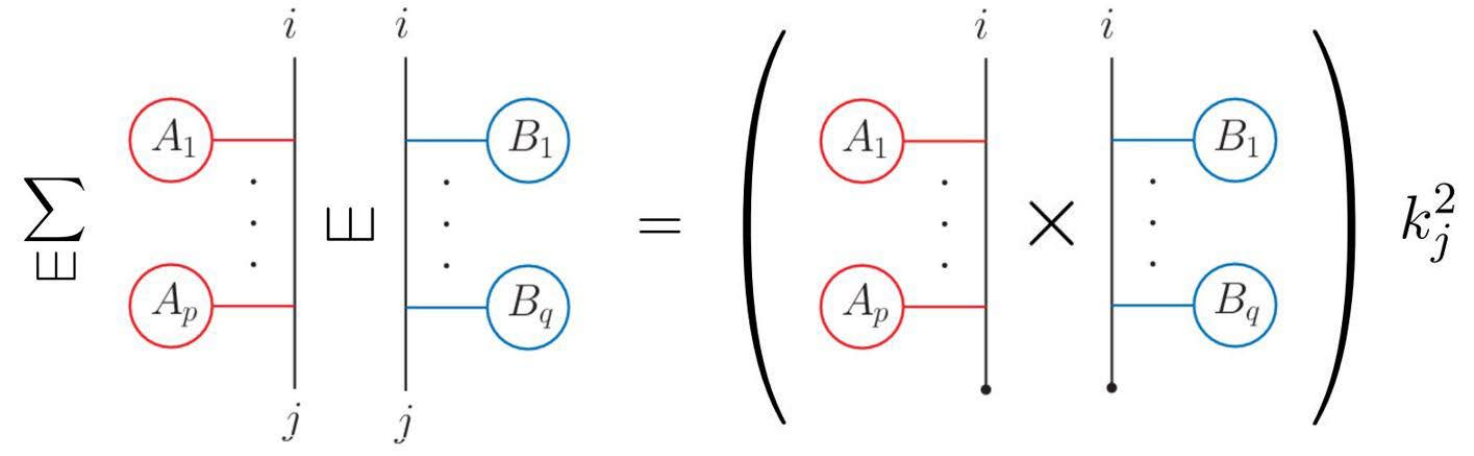} \\
  \caption{Diagrammatical interpretation for hidden zero. Each red circle or blue circle represents the BG current generated by the corresponding subset.}\label{Fig5}
\end{figure}

To interpret the behavior in \eref{zero-tree}, we first observe that in each connected Feynman diagram one can always find a line $L_{(i,j)}$ which connects external legs $i$ and $j$ together. Each tree-level Feynman diagram can be thought of as planting trees onto this line $L_{(i,j)}$.
Now consider dividing the ordered sets $\pmb A$ and $\pmb B$ into subsets, namely
\bea
\pmb A=\{A_1,A_2,\cdots,A_p\}\,,~~~~~~~~\pmb B=\{B_q,B_{q-1},\cdots,B_1\}\,.~~\label{division-zero-tree}
\eea
Each subset will generate a Berends-Giele (BG) current, and each BG current is connected to $L_{(i,j)}$ via a propagator, as shown in Fig.\ref{Fig5}.
Clearly, for $A_\a$ with $\a\in\{1,\cdots,p\}$, the propagator connecting the current ${\cal J}_{A_\a}$ to $L_{(i,j)}$ carries the momentum $k_{A_\a}$, which is the sum of all external momenta in the subset $A_\a$. Similarly, for $B_\b$ with $\b\in\{1,\cdots,q\}$, the propagator connecting ${\cal J}_{B_\b}$ to $L_{(i,j)}$ carries the momentum $k_{B_\b}$, defined as the sum of all external momenta in the subset $B_\b$. The zero kinematics in \eref{zero-tree} indicates that the momenta $k_{A_\a}$ and $k_{B_\b}$ satisfy the kinematic condition \eref{kine-condi-shuffle} for SFASL; consequently, the former are identified as $A$-lines and the latter as $B$-lines. Given a division of $\pmb A$ and $\pmb B$ in \eref{division-zero-tree}, the summation over all allowed Feynman diagrams yields, on the one hand, the BG currents generated by the subsets; on the other hand, it results in a summation over shuffles of these $A$-lines and $B$-lines, as illustrated on the l.h.s. of Fig.\ref{Fig5}. In what follows in this paper, we shall refer to the contribution (BG current) of each subset $A_\a$ or $B_\b$ as a block. This terminology is consistent with the representation of these subsets in Fig.\ref{Fig5} and subsequent figures.

However, we cannot yet directly apply the SFASL in \eref{fac-propa-gen}. The reason is, the line $L_{(i,\bullet)}$
in Fig.\ref{Fig1} ends at a cubic vertex $\bullet$, while the line $L_{(i,j)}$ ends at an eternal leg $j$. To use the SFASL in
\eref{fac-propa-gen}, we need to multiply the l.h.s. of Fig.\ref{Fig5} by $s_{iA_1\cdots A_p B_1\cdots B_q}/s_{iA_1\cdots A_p B_1\cdots B_q}$,
where the propagator $1/s_{iA_1\cdots A_p B_1\cdots B_q}$ compensates for the difference between the internal line and the external leg $j$.
Here
\bea
s_{iA_1\cdots A_p B_1\cdots B_q}=\Big(k_i+\sum_{\a=1}^p\,k_{A_\a}+\sum_{\b=1}^q\,k_{B_\b}\Big)^2=\Big(\sum_{m=j+1}^{j-1}\,k_m\Big)^2\,.
\eea
Then the factorization behavior in \eref{fac-propa-gen} implies the structure on the r.h.s. of Fig.\ref{Fig5}, i.e.,
\bea
\Big(\sum_{\shuffle(p,q)}\,\prod_{t=1}^{p+q}\,{1\over D_t^{(i,\bullet)}}\Big)\,s_{iA_1\cdots A_p B_1\cdots B_q}\,=\,\Big(\prod_{\a=1}^p\,{1\over s_{ia_1\cdots a_{\a}}}\Big)\,
\Big(\prod_{\b=1}^q\,{1\over s_{ib_1\cdots b_{\b}}}\Big)\,k_j^2\,,~~\label{zero-tree-interpre}
\eea
which is proportional to $k_j^2$ due to momentum conservation. Obviously, this $k_j^2$ will not be canceled by any $1/s_{ia_1\cdots a_\a}$ or $1/s_{ib_1\cdots b_\b}$ in two curly brackets; thus, the r.h.s. of Fig.\ref{Fig5} vanishes due to the on-shell condition $k_j^2=0$. This conclusion is valid for any division of $\pmb A$ and $\pmb B$, and therefore gives rise to the hidden zero in \eref{zero-tree}.

If one removes one leg from the zero kinematic in \eref{zero-tree}, the amplitude will exhibit $2$-split. Without loss of generality, let us choose this removed leg as $k\in\pmb B$. Then the explicit form of $2$-split is given as
\bea
{\cal A}_n(1,\cdots,n)\,&\to&\,{\cal J}_{n_1}(i,\pmb A,j,\kappa)\,\times\,{\cal J}_{n+3-n_1}(j,\pmb B(\kappa'),i)\,,
~~~~{\rm if}~k_{\hat{a}}\cdot k_{\hat{b}}=0\,,
~~~~{\rm for}~\forall\,\,\hat{a}\in\pmb A\,,~\hat{b}\in\pmb B\setminus k\,.~~\label{2split-tree}
\eea
In the above, ${\cal J}_{n_1}$ and ${\cal J}_{n+3-n_1}$ are two amputated currents, which carry off-shell external legs $\kappa$ and $\kappa'$,
respectively. The ordered set $\pmb B(\kappa')$ is obtained from $\pmb B$ by replacing $k$ with $\kappa'$\footnote{Here the terminology is slightly different from that in \cite{Cao:2024gln,Cao:2024qpp}. The set $B$ in \cite{Cao:2024gln,Cao:2024qpp} corresponds to $\pmb B\setminus k$ in this paper. We choose this way of expressing the kinematic conditions in order to make the connection between the zero condition and the $2$-split condition as manifest as possible.}.

\begin{figure}
  \centering
   \includegraphics[width=16cm]{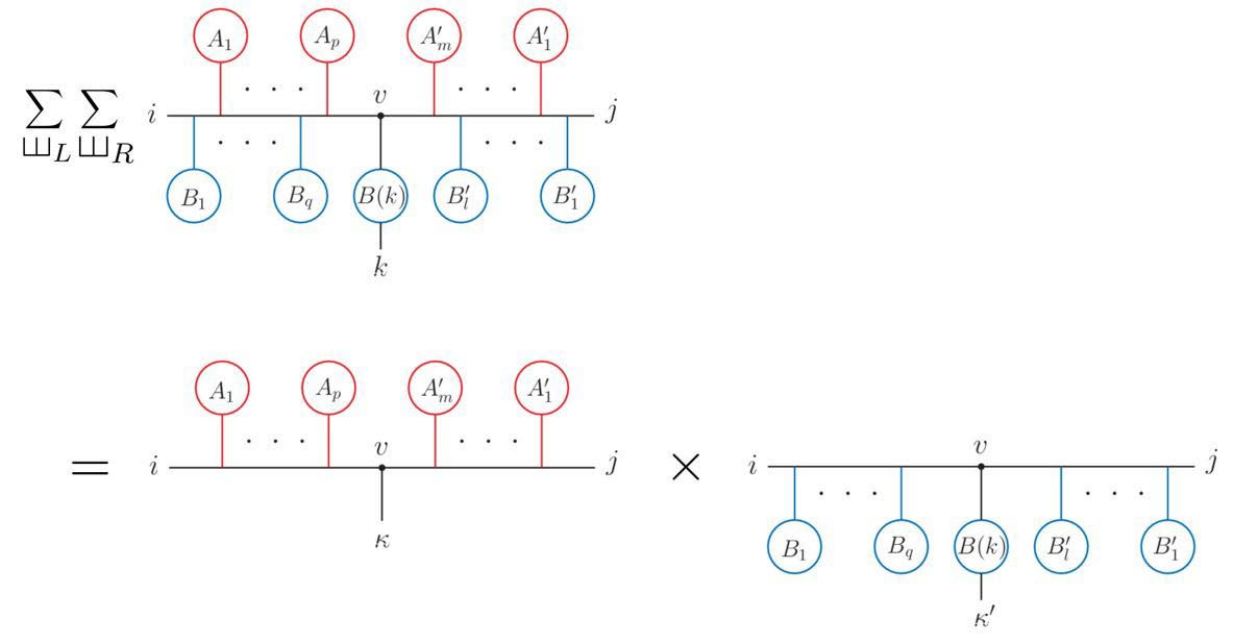} \\
  \caption{Diagrammatical interpretation for $2$-split. $\shuffle_L$ and $\shuffle_{R}$ in this graph correspond to $\shuffle_{(p,q)}$ and $\shuffle_{m,l}$ in \eref{two shuffles}, respectively. Although not shown explicitly, the first row of the figure actually contains all shuffles of $\{A_1,\cdots,A_p\}$ and $\{B_1,\cdots,B_q\}$ along the line $L_{(i,v)}$, as well as all shuffles of $\{A'_1,\cdots,A'_m\}$ and $\{B'_1,\cdots,B'_l\}$ along the line $L_{j,v}$.}\label{Fig6}
\end{figure}

To interpret this novel factorization behavior, we observe that in each Feynman diagram one can always find three lines $L_{i,v}$, $L_{j,v}$ and $L_{k,v}$, emanating from the external legs $i$, $j$ and $k$ respectively, meeting at a vertex $v$. Then we consider the division
\bea
\pmb A=\{A_1,\cdots,A_p,A'_m,\cdots,A'_1\}\,,~~~~\pmb B=\{B'_1,\cdots,B'_l,B(k),B_q,\cdots,B_1\}\,,~~\label{division-2split-tree}
\eea
where the special leg $k$ is included in $B(k)$. We further assume that blocks from $\{A_1,\cdots,A_p\}$ and $\{B_q,\cdots,B_1\}$ are planted onto $L_{(i,v)}$, blocks from $\{A'_m,\cdots,A'_1\}$ and $\{B'_1,\cdots,B'_l\}$ are planted onto $L_{(j,v)}$, while the block $B(k)$ is planted onto the vertex $v$, as illustrated in Fig.\ref{Fig6}. Using an argument similar to that in the hidden zero case, we see that blocks from $\{A_1,\cdots,A_p\}$ and $\{A'_m,\cdots,A'_1\}$ are connected to $L_{(i,v)}$ or $L_{(j,v)}$ via $A$-lines, while blocks from $\{B_q,\cdots,B_1\}$ and $\{B'_1,\cdots,B'_l\}$ are connected to $L_{(i,v)}$ or $L_{(j,v)}$ via $B$-lines. For a given division of $\pmb A$ and $\pmb B$ in \eref{division-2split-tree}, the summation over all allowed Feynman diagrams produces, on the one hand, the BG currents from the subsets; on the other hand, a summation over shuffles of $A$-lines and $B$-lines along $L_{i,v}$ or $L_{j,v}$.

Since $v$ is a vertex rather than an external leg, the SFASL in \eref{fac-propa-gen} can be directly applied to $L_{(i,v)}$
and $L_{(j,v)}$. Thus we obtain
\bea
\sum_{\shuffle(p,q)}\,\prod_{t=1}^{p+q}\,{1\over D_t^{(i,v)}}\,&=&\,\Big(\prod_{\a=1}^p\,{1\over s_{iA_1\cdots A_{\a}}}\Big)\,
\Big(\prod_{\b=1}^q\,{1\over s_{iB_1\cdots B_{\b}}}\Big)\,,\nn
\sum_{\shuffle(m,l)}\,\prod_{t=1}^{m+l}\,{1\over D_t^{(j,v)}}\,&=&\,\Big(\prod_{\a=1}^m\,{1\over s_{iA'_1\cdots A'_{\a}}}\Big)\,
\Big(\prod_{\b=1}^l\,{1\over s_{iB'_1\cdots B'_{\b}}}\Big)\,,~~\label{two shuffles}
\eea
which yields
\bea
&&\Big(\sum_{\shuffle(p,q)}\,\prod_{t=1}^{p+q}\,{1\over D_t^{(i,v)}}\Big)\,\Big(\sum_{\shuffle(m,l)}\,\prod_{t=1}^{m+l}\,{1\over D_t^{(j,v)}}\Big)\,
f(R)\nn
&=&\Big[\Big(\prod_{\a=1}^p\,{1\over s_{iA_1\cdots A_{\a}}}\Big)\,\Big(\prod_{\a=1}^m\,{1\over s_{iA'_1\cdots A'_{\a}}}\Big)\,f^A(R)\,\Big]\,\times\,\Big[\Big(\prod_{\b=1}^q\,{1\over s_{iB_1\cdots B_{\b}}}\Big)\,\Big(\prod_{\b=1}^l\,{1\over s_{iB'_1\cdots B'_{\b}}}\Big)\,f^B(R)\,\Big]\,,~~\label{meaning-fig6}
\eea
as illustrated in Fig.\ref{Fig6}.
In the above, $f(R)$, $f^A(R)$ and $f^B(R)$ are contributions from remaining parts of diagrams, and each of them can be expressed as a product of contributions from blocks shown in Fig.\ref{Fig6}:
\bea
f(R)&=&\Big(\prod_{\a=1}^p\,{\cal J}_{A_\a}\,{1\over s_{A_\a}}\Big)\,\Big(\prod_{\a=1}^m\,{\cal J}_{A'_\a}\,{1\over s_{A'_\a}}\Big)\,
\Big(\prod_{\b=1}^q\,{\cal J}_{B_\b}\,{1\over s_{B_\b}}\Big)\,\Big(\prod_{\b=1}^l\,{\cal J}_{B'_\b}\,{1\over s_{B'_\b}}\Big)\,\Big({\cal J}_{B(k)}\,{1\over s_{B(k)}}\Big)\,,\nn
f^A(R)&=&\Big(\prod_{\a=1}^p\,{\cal J}_{A_\a}\,{1\over s_{A_\a}}\Big)\,\Big(\prod_{\a=1}^m\,{\cal J}_{A'_\a}\,{1\over s_{A'_\a}}\Big)\,,\nn
f^B(R)&=&
\Big(\prod_{\b=1}^q\,{\cal J}_{B_\b}\,{1\over s_{B_\b}}\Big)\,\Big(\prod_{\b=1}^l\,{\cal J}_{B'_\b}\,{1\over s_{B'_\b}}\Big)\,\Big({\cal J}_{B(k)}\,{1\over s_{B(k)}}\Big)\,.~~~\label{fR}
\eea
Whether viewed from Fig.\ref{Fig6} or from the expression \eref{fR}, they satisfy the following relation,
\bea
f(R)\,=\,f^A(R)\,\times\,f^B(R)\,.~~\label{key-fR}
\eea

The factorization behavior described in \eref{meaning-fig6} and Fig.\ref{Fig6} is valid for any division of $\pmb A$ and $\pmb B$. Consequently, summing over all divisions denoted by ${\rm div}\pmb A$ and ${\rm div}\pmb B$ leads to
\bea
{\cal A}_n(1,\cdots,n)&=&\sum_{{\rm div}\pmb A}\,\sum_{{\rm div}\pmb B}\,\Big(\sum_{\shuffle(p,q)}\,\prod_{t=1}^{p+q}\,{1\over D_t^{(i,v)}}\Big)\,\Big(\sum_{\shuffle(m,l)}\,\prod_{t=1}^{m+l}\,{1\over D_t^{(j,v)}}\Big)\,
f(R)\nn
&\xrightarrow[]{k_{\hat{a}}\cdot k_{\hat{b}}=0}&\Big[\sum_{{\rm div}\pmb A}\Big(\prod_{\a=1}^p\,{1\over s_{iA_1\cdots A_{\a}}}\Big)\,\Big(\prod_{\a=1}^m\,{1\over s_{iA'_1\cdots A'_{\a}}}\Big)\,f^A(R)\,\Big]\nn
&&\times\,\Big[\sum_{{\rm div}\pmb B}\,\Big(\prod_{\b=1}^q\,{1\over s_{iB_1\cdots B_{\b}}}\Big)\,\Big(\prod_{\b=1}^l\,{1\over s_{iB'_1\cdots B'_{\b}}}\Big)\,f^B(R)\,\Big]\,,
\eea
which is precisely the $2$-split in \eref{2split-tree}, where
\bea
{\cal J}_{n_1}(i,\pmb A,j,\kappa)&=&\sum_{{\rm div}\pmb A}\Big(\prod_{\a=1}^p\,{1\over s_{iA_1\cdots A_{\a}}}\Big)\,\Big(\prod_{\a=1}^m\,{1\over s_{iA'_1\cdots A'_{\a}}}\Big)\,f^A(R)\,,\nn
{\cal J}_{n+3-n_1}(j,\pmb B(\kappa'),i)&=&\sum_{{\rm div}\pmb B}\,\Big(\prod_{\b=1}^q\,{1\over s_{iB_1\cdots B_{\b}}}\Big)\,\Big(\prod_{\b=1}^l\,{1\over s_{iB'_1\cdots B'_{\b}}}\Big)\,f^B(R)\,.
\eea

In two currents ${\cal J}_{n_1}$ and ${\cal J}_{n+3-n_1}$, momentum conservation forces $k_\kappa=k_{\pmb B}$, $k_{\kappa'}=k_k+k_{\pmb A}$,
where $k_{\pmb B}$ is the sum of momenta in $\pmb B$, namely, $k_{\pmb B}=\sum_{b=j+1}^{i-1}\,k_b$, and similarly $k_{\pmb A}=\sum_{a=i+1}^{j-1}\,k_a$. Since $f^B(R)$ contains ${\cal J}_{B(k)}/s_{B(k)}$ rather than ${\cal J}_{B(\kappa')}/s_{B(\kappa')}$, when defining ${\cal J}_{n+3-n_1}$, each propagator on the line $L_{(\kappa',v)}$ should be interpreted as,
\bea
{1\over(k_{\kappa'}+k_b)^2-(k_k+k_{\pmb A})^2}={1\over (k_k+k_b)^2}\,,
\eea
where the combined momentum $k_b$ consists of the momenta from a subset of $B(k)\setminus k$. That is, each propagator along $L_{(\kappa',v)}$ acquires a mass $m_{\kappa'}^2=(k_k+k_{\pmb A})^2$. By slightly extending the above idea, we can reinterpret the $2$-split \eref{2split-tree} as,
\bea
{\cal A}_n(1,\cdots,n)\,&\to&\,{\cal A}_{n_1}(i,\pmb A,j,\kappa)\,\times\,{\cal A}_{n+3-n_1}(j,\pmb B(\kappa'),i)\,,
\eea
where ${\cal A}_{n_1}$ and ${\cal A}_{n+3-n_1}$ are on-shell amplitudes. In ${\cal A}_{n_1}$, the external particle on the line $L_{(\kappa,v)}$ acquires a mass $m_\kappa^2=k_{\pmb B}^2$; in ${\cal A}_{n+3-n_1}$, each external/virtual particle on the line $L_{(\kappa',v)}$ acquires a mass $m_{\kappa'}^2=(k_k+k_{\pmb A})^2$.

%%%%%%%%%%%%%%%%%%%%%%%%%%%%%%%
\section{Hidden zero and $2$-split at $1$-loop level}
\label{sec-1loop}
%%%%%%%%%%%%%%%%%%%%%%%%%%%%%%%

In this section, through the mechanism of SFASL, we generalize the two properties---hidden zeros and $2$-split---from tree amplitudes to $1$-loop Feynman integrands. This generalization benefits from a feature of SFASL noted in section \ref{subsec-tree-proofshuffle}: it is a local property that depends solely on the interaction vertices on the line, and is completely independent of what the $A$-lines or $B$-lines are attached to. This feature allows us to carry out the generalization in a natural manner.

The $1$-loop hidden zero takes exactly the same form as at tree-level, whereas the $1$-loop $2$-split takes the form of a generalization of the tree-level $2$-split. More explicitly, the $1$-loop Feynman integrand is decomposed into two parts, each of which exhibits the $2$-split form, as shown in \eref{2split-1loop}. Among four products of such $2$-split, half of them contain a loop. We will elaborate on this point. Although we know which Feynman diagrams contribute to these loop-containing products, their physical interpretation remains unclear at present, as they differ from the standard Feynman integrands.

In subsection \ref{subsec-convention-1loop}, we give the kinematic conditions for hidden zeros and $2$-split of $1$-loop Feynman integrands, as well as the convention for choosing the loop momentum $\ell$. In subsection \ref{subsec-1loop-example}, we discuss the application of SFASL at the $1$-loop-level. In subsection \ref{subsec-1loop-zeroandsplit}, we use such generalized mechanism to establish the hidden zeros and $2$-split of $1$-loop Feynman integrands.

%%%%%%%%%%%%%%%%%%%%%%%%%%%%%%%
\subsection{Kinematic condition and loop-momentum convention}
\label{subsec-convention-1loop}
%%%%%%%%%%%%%%%%%%%%%%%%%%%%%%%

In this subsection, we specify the kinematic conditions that give rise to hidden zeros and $2$-split of $1$-loop Feynman integrands, as well as a convention for the assignment of the loop momentum.

The zero kinematics of an $n$-point, $1$-loop ${\rm Tr}(\phi^3)$ Feynman integrand is as follows.
As in the tree case, we pick a pair of external lines $(i,j)$, and separate the remaining external legs into $\pmb A=\{i+1,\cdots,j-1\}$
and $\pmb B=\{j+1,\cdots,i-1\}$. We then impose $k_{\hat{a}}\cdot k_{\hat{b}}=0$ for any $\hat{a}\in\pmb A$, $\hat{b}\in\pmb B$. In addition, we also
require the loop momentum to satisfy either $\ell\cdot k_{\hat{b}}=0$ or $\ell\cdot k_{\hat{a}}=0$. Without loss of generality, we choose the former condition in the rest of this paper. In summary, the kinematic condition for hidden zeros at $1$-loop level is given as
\bea
k_{\hat{a}}\cdot k_{\hat{b}}=0\,,~~~~\ell\cdot k_{\hat{b}}=0\,,~~~~{\rm for}~\forall\,\,\hat{a}\in\pmb A\,,~\hat{b}\in\pmb B\,.~~\label{kine-condi-1loop-0}
\eea

As in the tree-level case, the kinematic condition for $2$-split is closely related to the condition for hidden zero.
That is, the kinematic condition for $1$-loop $2$-split is obtained by removing $k\in\pmb B$ from \eref{kine-condi-1loop-0}:
\bea
k_{\hat{a}}\cdot k_{\hat{b}}=0\,,~~~~\ell\cdot k_{\hat{b}}=0\,,~~~~{\rm for}~\forall\,\,\hat{a}\in\pmb A\,,~\hat{b}\in\pmb B\setminus k\,.~~\label{kine-condi-1loop-split}
\eea
\begin{figure}
  \centering
   \includegraphics[width=6cm]{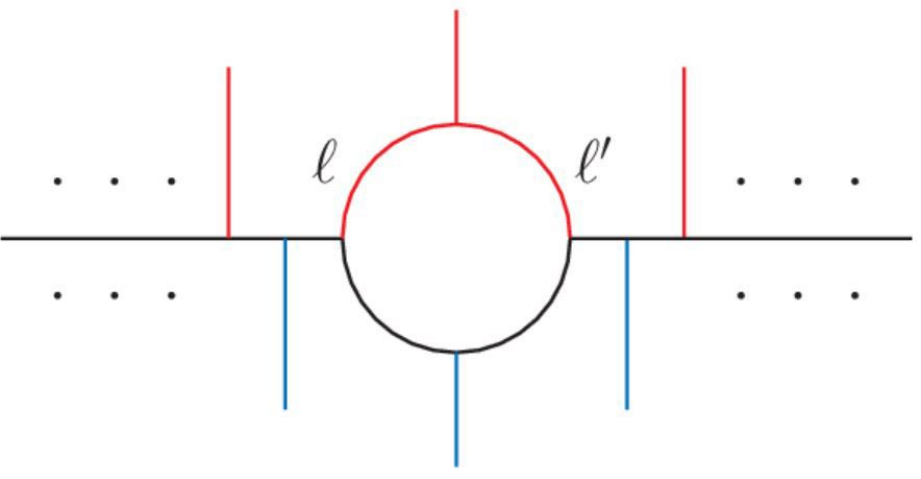} \\
  \caption{The convention for choosing the loop momentum: only $\ell$ and $\ell'$ are appropriate.}\label{Fig7}
\end{figure}

In order to derive the hidden zeros and $2$-split from the kinematic conditions \eref{kine-condi-1loop-0} and \eref{kine-condi-1loop-split}, a convention for the choice of the loop momentum is needed. This convention is illustrated in Fig.\ref{Fig7}.
In the tree case, $A$-lines and $B$-lines arising from $\pmb A$ and $\pmb B$ are attached solely to the line $L_{(i,j)}$. However,
in the $1$-loop case, if the loop is on the line $L_{(i,j)}$, such $A$-lines and $B$-lines can also be attached to the loop, as can be seen in Fig.\ref{Fig7}. For the case where the loop is on $L_{(i,j)}$, one can always divide the full loop into the $A$-side and $B$-side parts. For instance, in Fig.\ref{Fig7}, the loop is divided into upper (red) and lower (black) halves, which we refer to as the $A$-side part and the $B$-side part, respectively. We stipulate that when the loop lies on the line $L_{(i,j)}$, the loop momentum must be chosen from the $A$-side. For instance, in Fig.\ref{Fig7}, either $\ell$ or $\ell'$ are allowed. It is easy to see that under this stipulation, the kinematic condition \eref{kine-condi-1loop-0} or \eref{kine-condi-1loop-split} forces all propagators belong to the $A$-side part of the loop to be $A$-lines, since their momenta satisfy \eref{kine-condi-shuffle}. This is why the $A$-side loop is colored red in Fig.\ref{Fig7}. Allowing the loop momentum to be chosen from the $B$-side would lead to an inconsistency with the kinematic condition $\ell\cdot k_{\hat{b}}=0$.

In the following subsections, we will demonstrate how the kinematic conditions \eref{kine-condi-1loop-0} and \eref{kine-condi-1loop-split}, along with the convention for choosing the loop momentum, result in the hidden zeros and $2$-split of $1$-loop Feynman integrands.

%%%%%%%%%%%%%%%%%%%%%%%%%%%%%%%
\subsection{Generalization of shuffle factorization along special line}
\label{subsec-1loop-example}
%%%%%%%%%%%%%%%%%%%%%%%%%%%%%%%

In this subsection, we argue that the SFASL described by \eref{fac-propa-gen} and Fig.\ref{Fig1} can be applied to the case where there is a loop on the line.

\begin{figure}
  \centering
   \includegraphics[width=14cm]{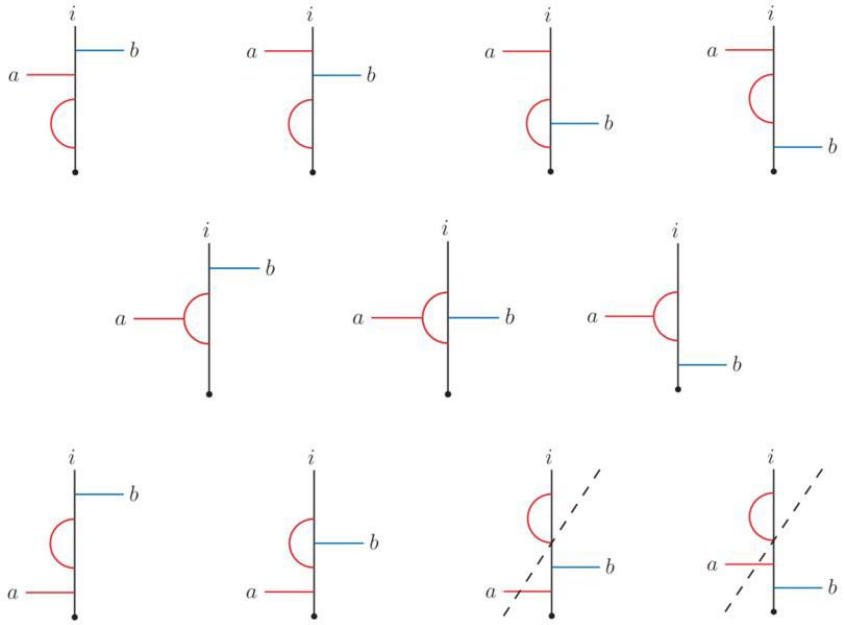} \\
  \caption{The effective diagrams for the example in section \ref{subsec-1loop-example}.}\label{Fig8}
\end{figure}

Consider the example illustrated in Fig.\ref{Fig8}: a loop lies on the line $L_{(i,\bullet)}$; there is one $A$-line, $a$, and one $B$-line, $b$, without distinguishing on-shell or off-shell. Note that when drawing diagrams in Fig.\ref{Fig8}, we have regarded the $B$-side part of loop in Fig.\ref{Fig7} as a part of $L_{(i,\bullet)}$. In this sense, the loop is formed by connecting $A$-lines onto $L_{(i,\bullet)}$.
This will be the convention for diagrammatic representation adopted in the remainder of this paper. When $L_{(i,\bullet)}$ is understood in this manner, we shall no longer refer to the $A$-side part of a loop and the $B$-side part of a loop. Instead, in the rest of this section, any reference to the $A$-side or the $B$- side shall denote the $A$-side or the $B$-side of the line $L_{(i,\bullet)}$ (as understood in the above sense), namely, the red-line side or the blue-line side of $L_{(i,\bullet)}$.

After dropping unphysical diagrams that correspond to scaleless integrands, we obtain the effective Feynman diagrams shown in Fig.\ref{Fig8}. The last two diagrams in the last line also contain massless bubbles which correspond to scaleless integrands, thus we have used dashed lines to indicate that they should be removed. The reason for putting them in the last line is to visualize the complete set of shuffle permutations generated by moving $b$.

\begin{figure}
  \centering
   \includegraphics[width=14cm]{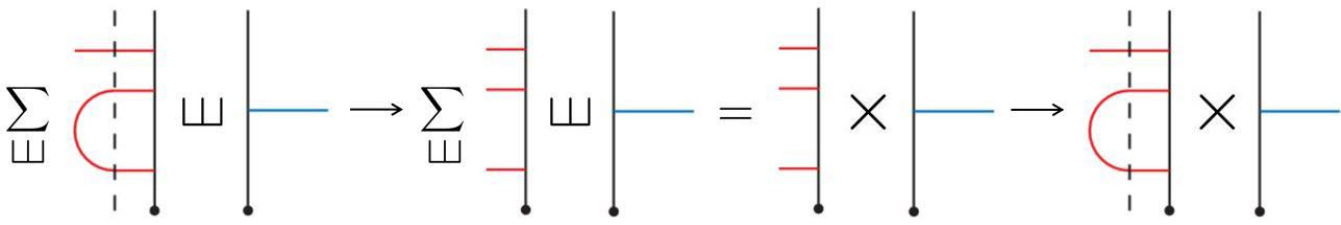} \\
  \caption{Treatment for the first line in Fig.\ref{Fig8}.}\label{Fig9}
\end{figure}

Let us begin with the four diagrams in the first line of Fig.\ref{Fig8}. In each of these diagrams, the loop is formed by attaching two ends of the red semicircle carrying $\ell$ to the black line $L_{(i,\bullet)}$. With a proper choice of loop momentum satisfying the convention provided at the end of subsection \ref{subsec-convention-1loop}, these diagrams can be evaluated as,
\bea
&&{1\over s_{ib}}\,{1\over s_{iab}}\,{1\over s_{iabl^-}}\,{1\over\ell^2}\,{1\over s_{iabl^-l^+}}+{1\over s_{ia}}\,{1\over s_{iab}}\,{1\over s_{iabl^-}}\,{1\over\ell^2}\,{1\over s_{iabl^-l^+}}\nn
&&+{1\over s_{ia}}\,{1\over s_{ial^-}}\,{1\over\ell^2}\,{1\over s_{iabl^-}}\,{1\over s_{iabl^-l^+}}\,+{1\over s_{ia}}\,{1\over s_{ial^-}}\,{1\over\ell^2}\,{1\over s_{ial^-l^+}}\,{1\over s_{iabl^-l^+}}\nn
&=&{1\over\ell^2}\,\Big(\sum_{\shuffle(3,1)}\,\prod_{t=1}^{3+1}\,{1\over D_t^{(i,\bullet)}}\Big)\nn
&=&{1\over\ell^2}\,\Big[\Big({1\over s_{ia}}\,{1\over s_{ial^-}}\,{1\over s_{ial^-l^+}}\,\Big)\,\times\,
{1\over s_{ib}}\Big]\nn
&=&\Big[{1\over\ell^2}\,\Big({1\over s_{ia}}\,{1\over s_{ial^-}}\,{1\over s_{ial^-l^+}}\,\Big)\Big]\,\times\,
\Big[{1\over s_{ib}}\Big]\,.~~\label{fig8-line1}
\eea
In the above, $l^-$ and $l^+$ represent two $A$-lines which carry the momenta $-\ell$ and $\ell$, respectively.
That is,
\bea
s_{ial^-}=(k_i+k_a-\ell)^2\,,~~~~s_{iabl^-}=(k_i+k_a+k_b-\ell)^2\,,~~~~s_{ial^-l^+}=s_{ia}\,,~~~~s_{iabl^-l^+}=s_{iab}\,.~~\label{rela-forl-l+}
\eea
Using these, one can directly verify that the first and second lines in \eref{fig8-line1} are equivalent to contributions from four diagrams computed via standard Feynman rules. The second step of \eref{fig8-line1} uses the SFASL expressed in \eref{fac-propa-gen}, with $(p,q)=(3,1)$. Note that the value of $p$ is $3$, since there are three $A$-lines: $a$, $l^-$ and $l^+$. The whole process is diagrammatically illustrated in Fig.\ref{Fig9}: we first cut off the components on the $A$-side,
then apply the SFASL in Fig.\ref{Fig1} to the remaining part, and ultimately glue the cut-off parts back into place. The three-step operation in Fig.\ref{Fig9} corresponds one-to-one with the three steps in \eref{fig8-line1}. The final result exhibits factorization into two objects: the left one contains a loop, while right one is a tree component. The loop in the left object does not involve any $B$-line. Therefore, the contribution from $\pmb B$ is decoupled from this loop.

In \eref{fig8-line1}, the applicability of the SFASL \eref{fac-propa-gen} reflects our claim in section \ref{subsec-tree-proofshuffle} that SFASL depends only on the interaction vertices on the specific line, and is independent of the structures of rest parts in the Feynman diagrams. More specifically, this local property does not distinguish between internal and external lines, nor does it distinguish whether the internal lines are connected to the BG currents or belong to loops.

\begin{figure}
  \centering
   \includegraphics[width=14cm]{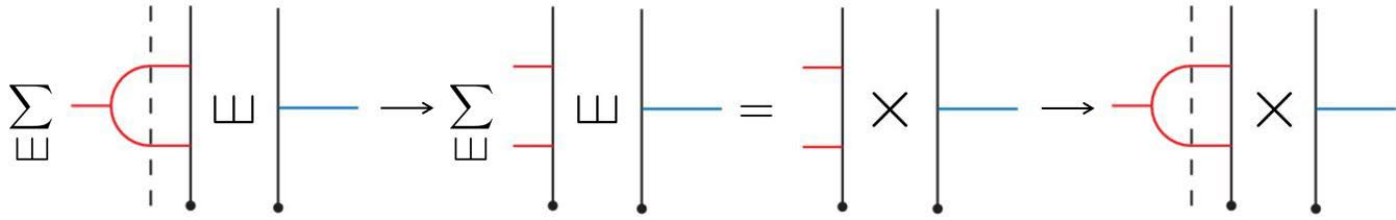} \\
  \caption{Treatment for the second line in Fig.\ref{Fig8}.}\label{Fig10}
\end{figure}

The three diagrams in the second line of Fig.\ref{Fig8} can be treated via a similar procedure. In each of these diagrams, the loop is formed by attaching two red lines that carry momenta $\ell$ and $\ell+k_a$, respectively, to the black line $L_{i,\bullet}$. With a proper choice of loop momentum, the computation can be carried out as follows,
\bea
&&{1\over s_{ib}}\,{1\over s_{ibl^-}}\,{1\over \ell^2}\,{1\over(\ell+k_a)^2}\,{1\over s_{ibl^-l^+}}
+{1\over s_{il^-}}\,{1\over \ell^2}\,{1\over(\ell+k_a)^2}\,{1\over s_{ibl^-}}\,{1\over s_{ibl^-l^+}}
+{1\over s_{il^-}}\,{1\over \ell^2}\,{1\over(\ell+k_a)^2}\,{1\over s_{il^-l^+}}\,{1\over s_{ibl^-l^+}}\nn
&=&{1\over \ell^2}\,{1\over(\ell+k_a)^2}\,\Big(\sum_{\shuffle(2,1)}\,\prod_{t=1}^{2+1}\,{1\over D_t^{(i,\bullet)}}\Big)\nn
&=&{1\over \ell^2}\,{1\over(\ell+k_a)^2}\,\Big[\Big({1\over s_{il^-}}\,{1\over s_{il^-l^+}}\Big)\,\times\,{1\over s_{ib}}\Big]\nn
&=&\Big[{1\over \ell^2}\,{1\over(\ell+k_a)^2}\,\Big({1\over s_{il^-}}\,{1\over s_{il^-l^+}}\Big)\Big]\,\times\,\Big[{1\over s_{ib}}\Big]\,.~~\label{fig8-line2}
\eea
In the above, $l^-$ and $l^+$ are two $A$-lines carry momenta $-\ell$ and $\ell+k_a$, respectively. Therefore,
\bea
s_{il^-}=(k_i-\ell)^2\,,~~~~s_{ibl^-}=(k_i+k_b-\ell)^2\,,~~~~s_{il^-l^+}=s_{ia}\,,~~~~s_{ibl^-l^+}=s_{iab}\,.
\eea
When applying the SFASL \eref{fac-propa-gen}, we have $p=2$, $q=1$, since vertices on the line $L_{i,\bullet}$ are sensitive to only two $A$-lines, $l^-$ and $l^+$. The process is illustrated in Fig.\ref{Fig10}: first remove the components on the $A$-side, then apply the SFASL in Fig.\ref{Fig1} to the remaining part, and finally reattach the removed components. Again, the result exhibits factorization into two objects: the left one contains a loop, while the right one is a tree component. As in the previous case, the contribution from $\pmb B$ is decoupled from the loop.

\begin{figure}
  \centering
   \includegraphics[width=14cm]{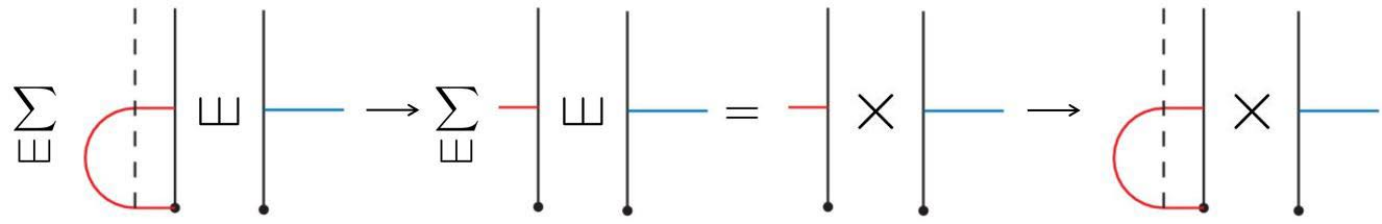} \\
  \caption{Treatment for the third line in Fig.\ref{Fig8}. We have omitted the red line connected to the vertex $\bullet$ in two middle graphs, since it does not enter the shuffle permutation. As shown, a massless bubble (on the left of $\times$) occurs.}\label{Fig11}
\end{figure}

Now we move to two diagrams in the last line of Fig.\ref{Fig8}. These two diagrams can be computed as,
\bea
&&{1\over s_{ib}}\,{1\over s_{ibl^-}}\,{1\over\ell^2}\,{1\over s_{ibl^-l^+}}\,{1\over s_{iabl^-l^+}}
+{1\over s_{il^-}}\,{1\over\ell^2}\,{1\over s_{ibl^-}}\,{1\over s_{ibl^-l^+}}\,{1\over s_{iabl^-l^+}}\nn
&=&{1\over\ell^2}\,{1\over s_{ibl^-l^+}}\,{1\over s_{iabl^-l^+}}\,\Big({1\over s_{ib}}\,{1\over s_{ibl^-}}
+{1\over s_{il^-}}\,{1\over s_{ibl^-}}\Big)\nn
&=&{1\over\ell^2}\,{1\over s_{ibl^-l^+}}\,{1\over s_{iabl^-l^+}}\,\Big({1\over s_{il^-}}\,\times\,{1\over s_{ib}}\Big)\nn
&=&\Big[{1\over\ell^2}\,{1\over 2\,\ell\cdot k_i}\Big]\,{1\over s^2_{ib}}\,{1\over s_{iab}}\,.~~\label{scaless-example}
\eea
In the above, two $A$-lines $l^-$ and $l^+$ carry momenta $-\ell$ and $\ell$, respectively. The second step uses the SFASL in \eref{fac-propa-gen}, with $p=q=1$; the third uses relations in \eref{rela-forl-l+}. The application of SFASL in this case differs slightly from the previous ones. Since two other diagrams are deleted in the last line of Fig.\ref{Fig8}, the set of shuffle permutations is incomplete. However, we can choose the end point of $L_{(i,\bullet)}$ as in Fig.\ref{Fig11} rather than in Fig,\ref{Fig8}, and apply the SFASL to such a new $L_{(i,\bullet)}$. The diagrammatic process is then the same as earlier, as can be seen in Fig.\ref{Fig11}.
We need not further consider the factorization structure of the result, as the loop-momentum-dependent part of the result is a scaleless integrand
which does not contribute to the physical amplitude.

The treatment of the example in Fig.\ref{Fig8} can be directly generalized to the general case. Since the SFASL described in Fig.\ref{Fig1} and \eref{fac-propa-gen} depends only on the local interactions on the line $L_{(i,\bullet)}$, when there is a loop on $L_{(i,\bullet)}$, one can always cut off the $A$-side components, retaining only the $A$-lines attached to $L_{(i,\bullet)}$, then apply the SFASL, and finally glue the removed parts back. Such manipulation results in a factorization pattern analogous to Fig.\ref{Fig1}, where the contribution from $\pmb B$ is decoupled from the loop. This constitutes the generalization of the SFASL in \eref{fac-propa-gen} to the $1$-loop level.

It is worth emphasizing that the above process of cutting off the $A$-side components and splicing them back merely rearranges the terms in the expression---for instance, by placing certain terms inside brackets or moving certain terms outside brackets, as we did in \eref{fig8-line1} and \eref{fig8-line2}. In other words, the cutting and splicing procedure does not have any substantive effect on the expression. This is because the parts being cut off or spliced back are completely independent of the momenta from the set $\pmb B$.

\begin{figure}
  \centering
   \includegraphics[width=12cm]{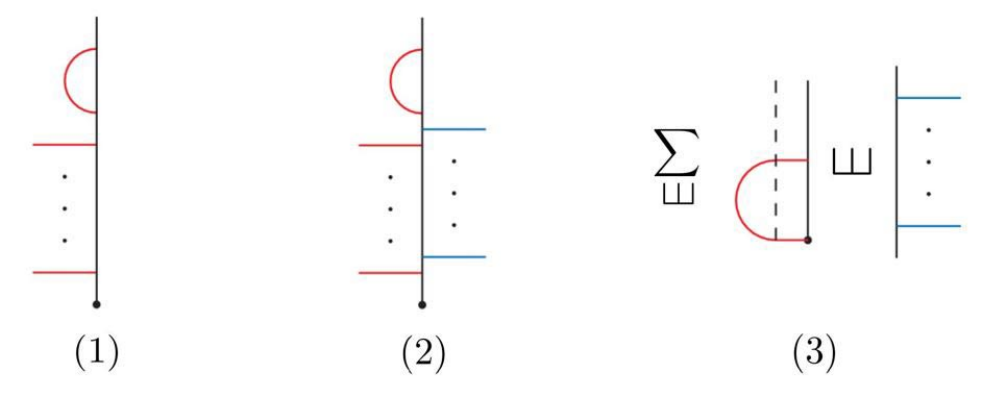} \\
  \caption{General procedure for handling massless bubbles. When the $A$-side configuration is given as graph $(1)$, the diagrams containing massless bubbles, such as the one in graph $(2)$, will be dropped. The treatment for this case is shown in graph $(3)$.}\label{scaleless}
\end{figure}

The only subtlety concerns the case in the last line of Fig.\ref{Fig8}. In general, when the $A$-side configuration participating in the shuffle permutation is as shown in graph $(1)$ of Fig.\ref{scaleless}, we always drop unphysical diagrams which contain the massless bubble, for instance the diagram in graph $(2)$. Consequently, the shuffle permutation becomes incomplete. However, we can always re-choose the endpoint of $L_{(i,\bullet)}$ as in graph $(3)$, just as we did in \eref{scaless-example}, and then apply the SFASL to this newly chosen $L_{(i,\bullet)}$. The above procedure again yields the scaleless integrand appearing in \eref{scaless-example}, due to the massless bubble occurring in graph $(3)$, and therefore such diagrams have no physical contribution and can be directly discarded. The above phenomenon is general, and the reason is simple. If some diagrams in the complete set of shuffle permutations contain a component that corresponds to a scaleless integral, then the left part involved in the shuffle permutation must also contain that same component. Consequently, it will appear on the l.h.s. in the final factorization formula.

One might raise a concern about the above procedure of discarding scaleless terms, on the grounds that these integrands are obtained under a constraint $\ell\cdot k_{\hat{b}}=0$ imposed on the loop momentum, whereas a scaleless integral vanishes only in the sense of integration, and integration necessarily means breaking the constraint on the loop momentum. The answer is that, when studying a $1$-loop integrand, we may subtract the scaleless terms we wish to discard from the integrand in advance, before imposing the kinematic constraint, thereby obtaining a physically equivalent new integrand. In this way, after the kinematic constraint is imposed, the resulting scaleless terms cancel against the pre-subtracted terms, so that the final result contains no scaleless contributions. Based on the above argument, we can be confident that discarding these scaleless terms is justified.

%%%%%%%%%%%%%%%%%%%%%%%%%%%%%%%
\subsection{Hidden zero and $2$-split of $1$-loop Feynman integrand}
\label{subsec-1loop-zeroandsplit}
%%%%%%%%%%%%%%%%%%%%%%%%%%%%%%%

Now we are ready to extend hidden zeros and $2$-split to the $1$-loop Feynman integrands.

It is evident that when the loop is not on the line $L_{(i,j)}$, all the arguments in section \ref{subsec-tree-zerosplit} carry over to the present $1$-loop case. Using the generalized SFASL discussed in the previous subsection \ref{subsec-1loop-example}, the discussion in section \ref{subsec-tree-zerosplit} can be repeated for the case where the loop lies on $L_{(i,j)}$ as well. The only distinction is that at tree-level, for a given division of the ordered sets
$\pmb A$ and $\pmb B$, it is enough to apply the SFASL. When a loop is present on the line $L_{(i,j)}$,
however, one must additionally specify the configuration on the $A$-side, beyond merely dividing $\pmb A$ and $\pmb B$. Only then can the position and momentum of each $A$-line be completely determined after the components on the $A$-side are cut off.

\begin{figure}
  \centering
   \includegraphics[width=12cm]{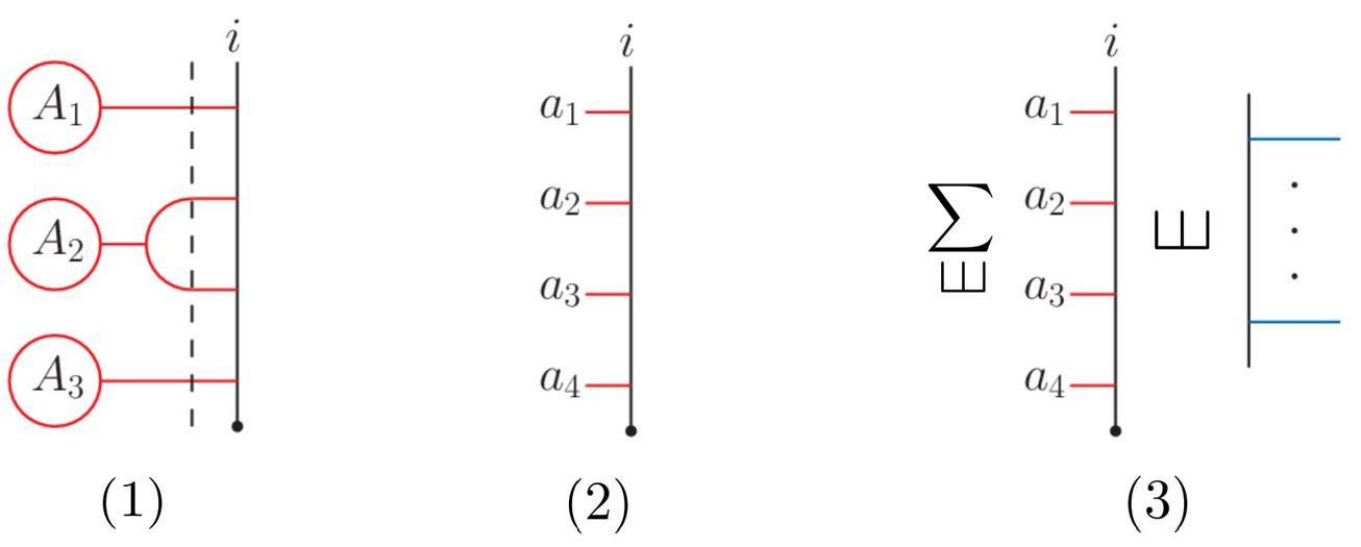} \\
  \caption{An example of ``$A$-side configuration".}\label{configuration-A}
\end{figure}

To be more precise, consider the example in Fig.\ref{configuration-A}. By ``configuration on the $A$-side", we mean the following: blocks $A_1$ and $A_3$ are attached to $L_{(i,\bullet)}$, block $A_2$ is attached to the loop, and the two ends of the semicircle are attached to $L_{(i,\bullet)}$ between $A_1$
and $A_3$, as illustrated in graph $(1)$ of Fig.\ref{configuration-A}. With this configuration clarified, the $A$-lines resulting from cutting off $A$-side components are determined as $a_1$, $a_2$, $a_3$, $a_4$, which carry momenta $k_{A_1}$, $-\ell$, $\ell+k_{A_2}$, $k_{A_3}$, respectively (see graph $(2)$ in Fig.\ref{configuration-A}). One can then consider the summation over shuffle permutations in graph $(3)$ and apply the SFASL in \eref{fac-propa-gen}.

Therefore, when the loop lies on the line $L_{(i,j)}$, the summation over Feynman diagrams can be decomposed into three levels: summation over all divisions of $\pmb A$ and $\pmb B$; summation over all configurations on the
$A$-side for a given division; and summation over all shuffle permutations for a given division and a given
$A$-side configuration. As clarified in subsection \ref{subsec-1loop-example}, when summing over shuffle permutations, one can safely apply the SFASL.

Using the treatment described above, we first obtain the $1$-loop extension of the structure in Fig.\eref{Fig5} and \eref{zero-tree-interpre}.
In the extended version of Fig.\ref{Fig5}, one of two parts in the curly brackets contains a loop. However, the factor outside the curly brackets remains $k_j^2$.
Therefore, the kinematic condition \eref{kine-condi-1loop-0} leads to the hidden zero of $1$-loop Feynman integrands, namely,
\bea
{\cal I}^{1-{\rm loop}}_n(1,\cdots,n)\,&\xrightarrow[]{\eref{kine-condi-1loop-0}}&\,0\,,~~\label{zero-1loop}
\eea
due to the on-shell property $k_j^2=0$. Of course, one must discard the scaleless integrals according to the procedure discussed at the end of the preceding subsection; only under this condition does the above hidden zero hold.

\begin{figure}
  \centering
   \includegraphics[width=16cm]{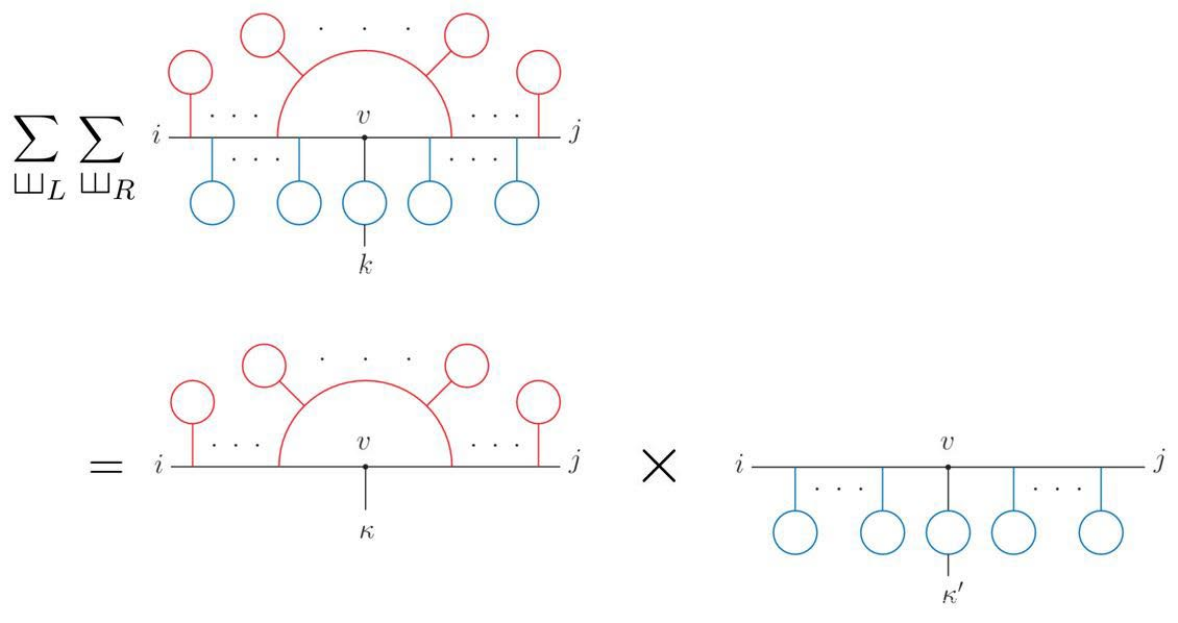} \\
  \caption{Generalization of the factorization pattern in Fig.\ref{Fig6} at $1$-loop-level.}\label{Fig12}
\end{figure}

The factorization in Fig.\ref{Fig6} and \eref{meaning-fig6} for a given division of $\pmb A$ and $\pmb B$ can also be directly generalized to the $1$-loop level. Let us consider a most nontrivial case. In this case, as illustrated in Fig.\ref{Fig12}, the loop lies on $L_{(i,j)}$, and the vertex $v$---which connects $L_{(i,v)}$, $L_{(j,v)}$, and $L_{(k,v)}$---lies on the loop. Assume that the $A$-side configuration is given. After cutting off $A$-side components, Fig.\ref{Fig12} becomes entirely equivalent to Fig.\ref{Fig6}. This indicates that the SFASL can be applied without distinction, thereby ensuring the factorizations exhibited in \eref{two shuffles}. On the other hand, we need to examine another foundation of the factorization pattern in Fig.\ref{Fig6}: the factorization property in \eref{key-fR}.
When the loop lies on $L_{(i,j)}$, each of two functions $f(R)$ and $f^A(R)$ in \eref{fR} contains a new factor (with a proper choice of loop momentum),
\bea
{1\over\ell^2}\,\prod_{\a=0}^{r}\,{1\over (\ell+k_{A_l\cdots A_{l+\a}})^2}\,,~~~~{\rm where}~k_{A_l\cdots A_{l+\a}}=k_{A_l}+k_{A_{l+1}}+\cdots +k_{A_{l+\a}}\,.~~\label{loop-fator}
\eea
This factor corresponds to the configuration that the blocks from $\{A_l,\cdots,A_{l+r}\}$ are attached to the loop rather than directly to $L_{(i,j)}$.
As two examples, this factor is $1/\ell^2$ in \eref{fig8-line1}, and ${1\over\ell^2}{1\over (\ell+k_a)^2}$ in \eref{fig8-line2}.
Clearly, at the $1$-loop level, $f(R)$, $f^A(R)$ and $f^B(R)$ continue to satisfy the factorization property \eref{key-fR}, since the new factor \eref{loop-fator} does not receive any contribution from $\pmb B$. This property, together with the SFASL along $L_{(i,v)}$ and $L_{(j,v)}$, guarantees the validity of the factorization depicted in Fig.\ref{Fig12}.

\begin{figure}
  \centering
   \includegraphics[width=9cm]{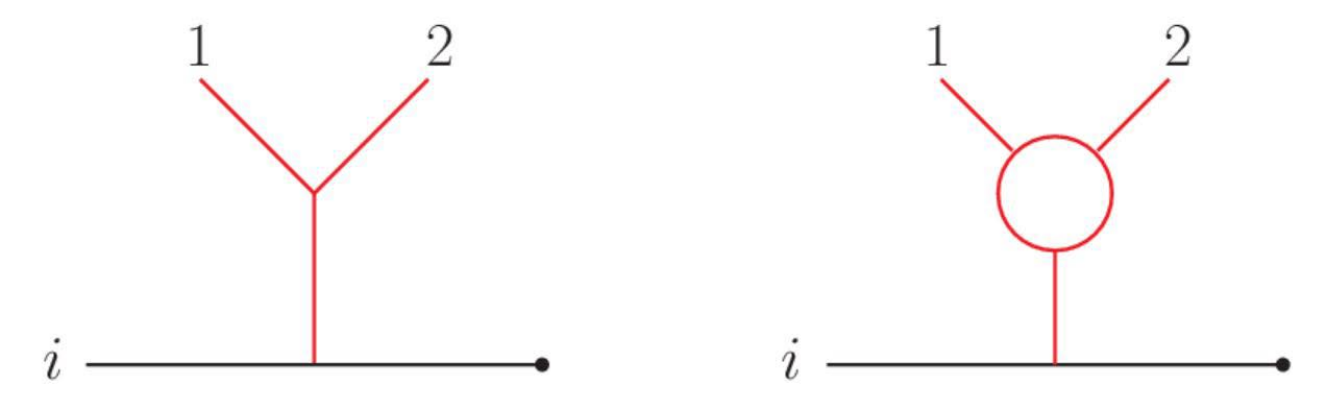} \\
  \caption{A simple example of $A$-replacement: the tree-level block on the left should be replaced by the $1$-loop block on the right. }\label{A-replace}
\end{figure}

Overall, two reasons jointly ensure that, for a given division of $\pmb A$ and $\pmb B$, the factorization pattern \eref{meaning-fig6} holds at $1$-loop level, with neither reason being dispensable. The first reason is that, according to the discussion in subsection \ref{subsec-1loop-example}, the SFASL in \eref{fac-propa-gen} holds universally, and therefore, the factorizations in \eref{two shuffles} follow without exception. The second reason is that another key relation \eref{key-fR} can be established, which needs to be verified in two separate cases. The first case is, (i) the loop lies on $L_{(i,j)}$, and (ii) the $A$-side configuration is given. Given that we introduce the new factor \eref{loop-fator} for $f^A(R)$ without affecting the key relation \eref{key-fR}, the factorization behavior in \eref{meaning-fig6} is guaranteed to hold in this case. The second case is, the loop does not lie on $L_{(i,j)}$. In this case, one needs to replace a certain tree-level BG current
${\cal J}$ in \eref{fR} with its counterpart containing a loop. For instance, consider a specific subset $A_1$ in a certain division of $\pmb A$ which is given as $A_1=\{1,2\}$, as illustrated in Fig.\ref{A-replace}. At tree-level, the BG current contributed by $A_1$ is just the constant ${\cal J}_{A_1}=1$, but at $1$-loop, suppose the loop is contributed by this block, then ${\cal J}_{A_1}=1$ should be replaced by ${1\over\ell^2}{1\over(\ell+k_1)^2}{1\over(\ell+k_1+k_2)^2}$ (with scaleless terms dropped). We refer to such a replacement as an $A$-replacement or a $B$-replacement, depending on whether it occurs on the $A$-side or the $B$-side. An $A$-replacement or a $B$-replacement likewise does not affect the correctness of \eref{key-fR}, thus the factorization pattern in \eref{meaning-fig6} is again ensured.

Summing over all divisions of $\pmb A$ and $\pmb B$, and, for each division, over all $A$-side configurations, $A$-replacements, and $B$-replacements, we get
\bea
{\cal I}^{1-{\rm loop}}_n(1,\cdots,n)\,&=&\,\sum_{{\rm div}\pmb A}\,\sum_{{\rm div}\pmb B}\,\sum_{N_{(i,j)}}\sum_{N_A}\sum_{N_B}\,\delta(N_{(i,j)}+N_A+N_B-1)\,\sum_{\rm AC}\sum_{\rm AR}\sum_{\rm BR}\nn
&&\Big(\sum_{\shuffle(p,q)}\,\prod_{t=1}^{p+q}\,{1\over D_t^{(i,v)}}\Big)\,\Big(\sum_{\shuffle(m,l)}\,\prod_{t=1}^{m+l}\,{1\over D_t^{(j,v)}}\Big)\,
f(R)\nn
&\xrightarrow[]{\eref{kine-condi-1loop-split}}&\,\Big[\sum_{{\rm div}\pmb A}\,\sum_{N_{(i,j)}}\sum_{N_A}\,\delta(N_{(i,j)}+N_A-1)\,\sum_{\rm AC}\sum_{\rm AR}\,\Big(\prod_{\a=1}^p\,{1\over s_{iA_1\cdots A_{\a}}}\Big)\,\Big(\prod_{\a=1}^m\,{1\over s_{iA'_1\cdots A'_{\a}}}\Big)\,f^A(R)\,\Big]\nn
&&\times\,\Big[\sum_{{\rm div}\pmb B}\,\Big(\prod_{\b=1}^q\,{1\over s_{iB_1\cdots B_{\b}}}\Big)\,\Big(\prod_{\b=1}^l\,{1\over s_{iB'_1\cdots B'_{\b}}}\Big)\,f^B(R)\,\Big]\nn
&&+\Big[\sum_{{\rm div}\pmb A}\,\Big(\prod_{\a=1}^p\,{1\over s_{iA_1\cdots A_{\a}}}\Big)\,\Big(\prod_{\a=1}^m\,{1\over s_{iA'_1\cdots A'_{\a}}}\Big)\,f^A(R)\,\Big]\nn
&&\times\,\Big[\sum_{{\rm div}\pmb B}\,\sum_{\rm BR}\,\Big(\prod_{\b=1}^q\,{1\over s_{iB_1\cdots B_{\b}}}\Big)\,\Big(\prod_{\b=1}^l\,{1\over s_{iB'_1\cdots B'_{\b}}}\Big)\,f^B(R)\,\Big]\,,~~~~\label{2split-1loop-detail}
\eea
where $\sum_{\rm AC}$, $\sum_{\rm AR}$ and $\sum_{\rm BR}$ are summations for $A$-side configurations, $A$-replacements and $B$-replacements, respectively.
$N_{(i,j)}$, $N_A$ and $N_B$ are the numbers of loops on the line $L_{(i,j)}$, $A$-side and $B$-side, respectively, satisfying $N_{(i,j)}+N_A+N_B=1$. Using the notations $N_{(i,j)}$, $N_A$ and $N_B$
at the $1$-loop-level may seem redundant, but this formulation facilitates the generalization to higher loops in the next section.

The result in \eref{2split-1loop-detail} is precisely the generalized $2$-split at the $1$-loop-level, which can be recast as,
\bea
{\cal I}^{1-{\rm loop}}_n(1,\cdots,n)\,&\xrightarrow[]{\eref{kine-condi-1loop-split}}&\,\W{\cal I}^{1-{\rm loop}}_{n_1}(i,\pmb A,j,\kappa)\,\times\,{\cal J}^{0-{\rm loop}}_{n+3-n_1}(j,\pmb B(\kappa'),i)\nn
&&+{\cal J}^{0-{\rm loop}}_{n_1}(i,\pmb A,j,\kappa)\,\times\,\W{\cal I}^{1-{\rm loop}}_{n+3-n_1}(j,\pmb B(\kappa'),i)\,,~~\label{2split-1loop}
\eea
where
\bea
\W{\cal I}^{1-{\rm loop}}_{n_1}(i,\pmb A,j,\kappa)&=&\sum_{{\rm div}\pmb A}\,\sum_{N_{(i,j)}}\sum_{N_A}\,\delta(N_{(i,j)}+N_A-1)\,\sum_{\rm AC}\sum_{\rm AR}\,\Big(\prod_{\a=1}^p\,{1\over s_{iA_1\cdots A_{\a}}}\Big)\,\Big(\prod_{\a=1}^m\,{1\over s_{iA'_1\cdots A'_{\a}}}\Big)\,f^A(R)\,,\nn
{\cal J}^{0-{\rm loop}}_{n+3-n_1}(j,\pmb B(\kappa'),i)&=&\sum_{{\rm div}\pmb B}\,\Big(\prod_{\b=1}^q\,{1\over s_{iB_1\cdots B_{\b}}}\Big)\,\Big(\prod_{\b=1}^l\,{1\over s_{iB'_1\cdots B'_{\b}}}\Big)\,f^B(R)\,,\nn
{\cal J}^{0-{\rm loop}}_{n_1}(i,\pmb A,j,\kappa)&=&\sum_{{\rm div}\pmb A}\,\Big(\prod_{\a=1}^p\,{1\over s_{iA_1\cdots A_{\a}}}\Big)\,\Big(\prod_{\a=1}^m\,{1\over s_{iA'_1\cdots A'_{\a}}}\Big)\,f^A(R)\,,\nn
\W{\cal I}^{1-{\rm loop}}_{n+3-n_1}(j,\pmb B(\kappa'),i)&=&\sum_{{\rm div}\pmb B}\,\sum_{\rm BR}\,\Big(\prod_{\b=1}^q\,{1\over s_{iB_1\cdots B_{\b}}}\Big)\,\Big(\prod_{\b=1}^l\,{1\over s_{iB'_1\cdots B'_{\b}}}\Big)\,f^B(R)\,.~~\label{4J}
\eea
In the above, ${\cal J}^{0-{\rm loop}}_{n+3-n_1}$ and ${\cal J}^{0-{\rm loop}}_{n_1}$ are exactly the tree-level currents.
It is worth emphasizing that this $2$-split formula holds only when the scaleless integrals are discarded in the manner discussed at the end of the previous subsection.

The first line on the r.h.s. of \eref{2split-1loop} arises from diagrams in which the loop is on the $A$-side or on $L_{(i,j)}$, due to the summation over $A$-side configurations and $A$-replacements in the expression of $\W{\cal I}^{1-{\rm loop}}_{n_1}$ in \eref{4J}. The loop-level object $\W{\cal I}^{1-{\rm loop}}_{n_1}$
is denoted as $\W{\cal I}$ rather than ${\cal I}$, since there are two reasons why we are uncertain whether it can be called a Feynman integrand. First, it carries an off-shell external leg $\kappa$. Second, although it contains contributions from most cubic diagrams compatible with the set of external legs $\{i,\cdots,j,\kappa\}$, it does not receive contributions from those where the loop lies on $L_{(\kappa,v)}$. Since $\kappa$ is off-shell/massive, these missing terms are no longer scaleless\footnote{One cannot attempt to use the constraint $\ell\cdot k_{\hat{b}}=0$ to turn the integrands corresponding to such diagrams into scaleless integrands. The situation differs from that at the end of the previous subsection: we may subtract certain scaleless terms from the original integrand to cancel the scaleless terms obtained under the kinematic constraint; however, we cannot transform a non-scaleless integrand into a scaleless one by imposing the constraint, as this would alter the space over which the integration is performed.}.

\begin{figure}
  \centering
   \includegraphics[width=6cm]{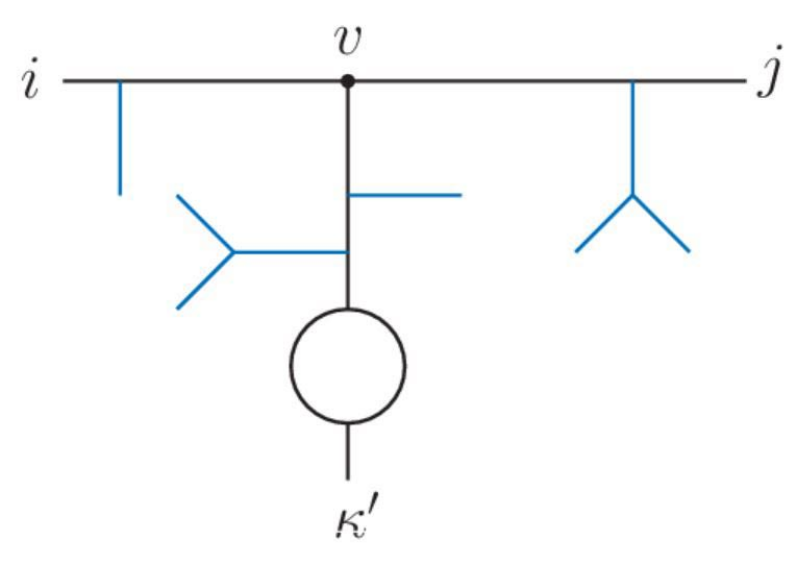} \\
  \caption{An example of diagrams containing a $\kappa'$-bubble.}\label{k-bubble}
\end{figure}

The second line on the r.h.s. of \eref{2split-1loop} originates from diagrams in which the loop is on the $B$-side, due to the summation over $B$-replacements in the expression of $\W{\cal I}^{1-{\rm loop}}_{n+3-n_1}$ in \eref{4J}. For similar reasons, it is not clear whether $\W{\cal I}^{1-{\rm loop}}_{n+3-n_1}$ may be termed a Feynman integrand: it carries an off-shell external leg $\kappa'$, and it excludes contributions from diagrams where the loop lies on $L_{(i,j)}$, as well as diagrams that contain one component which can be called a $\kappa'$-bubble. A $\kappa'$-bubble refers to a bubble to which the external leg $\kappa'$ is directly attached. Fig.\ref{k-bubble} provides an example containing a $\kappa'$-bubble. The external leg $\kappa'$ originates from the external leg $k$ in the original integrand ${\cal I}_n^{1-{\rm loop}}$. In the original integrand ${\cal I}_n^{1-{\rm loop}}$, the external leg $k$ is a massless on-shell particle; consequently, all diagrams containing a $k$-bubble are discarded due to the corresponding scaleless integrals. This causes that the resulting $\W{\cal I}^{1-{\rm loop}}_{n+3-n_1}$ does not include diagrams containing a $\kappa'$-bubble.

There is a third notable difference between $\W{\cal I}^{1-{\rm loop}}_{n+3-n_1}$ and a standard Feynman integrand. The $2$-split is derived based on the kinematic condition \eref{kine-condi-1loop-split}, and the presence of
$\ell\cdot k_{\hat{b}}=0$ in this kinematic condition results in
$\W{\cal I}^{1-{\rm loop}}_{n+3-n_1}$ containing no $\ell\cdot k_{B^{\rm sub}}$ in any denominator (since the kinematic condition does not involve constraints such as $\ell\cdot k_{\hat{a}}=0$, no similar issue arises in $\W{\cal I}^{1-{\rm loop}}_{n_1}$). Here $B^{\rm sub}$ stands for a subset of $\pmb B\setminus k$, and $k_{B^{\rm sub}}$ is the total momentum carried by all external legs in $B^{\rm sub}$. This somewhat peculiar structure renders the physical interpretation of $\W{\cal I}^{1-{\rm loop}}_{n+3-n_1}$ even more obscure.

%%%%%%%%%%%%%%%%%%%%%%%%%%%%%%%%
\section{Extension to higher loops}
\label{sec-higherloop}
%%%%%%%%%%%%%%%%%%%%%%%%%%%%%%%%

In this section, using a method completely analogous to that in the previous section, we generalize the hidden zeros and $2$-split to Feynman integrands at higher loops. The connection between the kinematic condition for hidden zeros and the kinematic condition for $2$-split at higher loops is perfectly aligned with that at tree-level and $1$-loop-level. The resulting formula of $2$-split of multi-loop Feynman integrands further manifests itself as a generalization of the tree-level $2$-split: the $L$-loop Feynman integrand decomposes into $L+1$ parts, each of which exhibits the $2$-split form.

In subsection \ref{subsec-higherloop-shuffle}, we show the application of SFASL at higher loops. Then, in subsection \ref{subsec-higherloop-zeroandsplit}, we use this mechanism to establish the hidden zeros and $2$-split of multi-loop Feynman integrands.

%%%%%%%%%%%%%%%%%%%%%%%%%%%%%%%
\subsection{Factorization along special line}
\label{subsec-higherloop-shuffle}
%%%%%%%%%%%%%%%%%%%%%%%%%%%%%%%

%
\begin{figure}
  \centering
   \includegraphics[width=8cm]{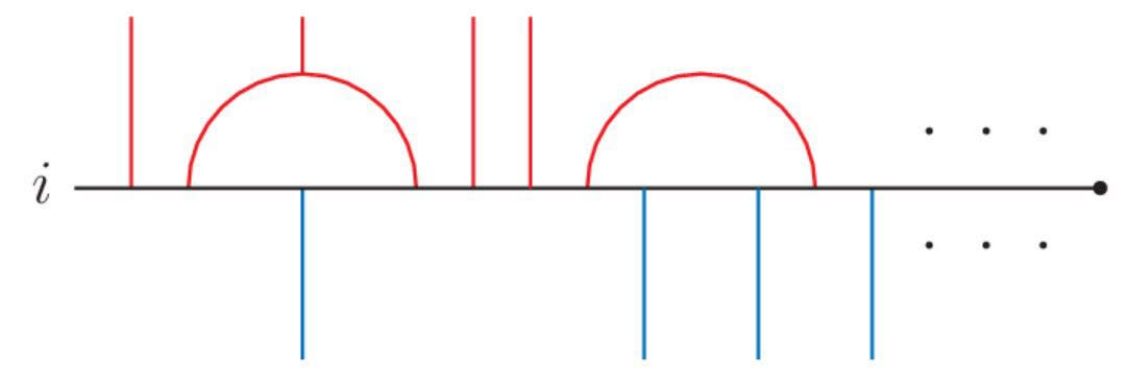} \\
  \caption{An example where the loops on the line $L_{(i,\bullet)}$ are independent.}\label{independent-loops}
\end{figure}

Extending the SFASL to the multi-loop case is quite straightforward. For an $n$-point, $L$-loop Feynman integrand, we first extend the kinematic conditions of hidden zeros and $2$-split as
\bea
k_{\hat{a}}\cdot k_{\hat{b}}=0\,,~~~~\ell_m\cdot k_{\hat{b}}=0\,,~~~~{\rm for}~\forall\,\,\hat{a}\in\pmb A\,,~\hat{b}\in\pmb B\,,~~m\in\{1,\cdots,L\}\,,~~\label{kine-condi-Lloop-0}
\eea
and
\bea
k_{\hat{a}}\cdot k_{\hat{b}}=0\,,~~~~\ell_m\cdot k_{\hat{b}}=0\,,~~~~{\rm for}~\forall\,\,\hat{a}\in\pmb A\,,~\hat{b}\in\pmb B\setminus k\,,~~m\in\{1,\cdots,L\}\,,~~\label{kine-condi-Lloop-split}
\eea
respectively, where $\ell_m$ is the loop momentum of the $m^{\rm th}$ loop. As in the $1$-loop case, when a loop lies on $L_{(i,\bullet)}$,
we stipulate that the corresponding loop momentum must be chosen from the $A$-side part of the loop, and the part of the loop that is not composed of $A$-lines is regarded as part of $L_{(i,\bullet)}$. Again, after $L_{(i,\bullet)}$ is understood in this way, the kinematic condition \eref{kine-condi-Lloop-0} or \eref{kine-condi-Lloop-split} implies that the propagators on loops attached to $L_{(i,\bullet)}$ are $A$-lines. Under such conditions and conventions, it is evident that if the loops on $L_{(i,\bullet)}$
are independent of each other, as illustrated for example in Fig. \ref{independent-loops}, then all discussions in section \ref{subsec-1loop-example} apply to the multi-loop case as well. Therefore, we need only concern ourselves with situations where loops are entangled with each other by sharing propagators.

\begin{figure}
  \centering
   \includegraphics[width=14cm]{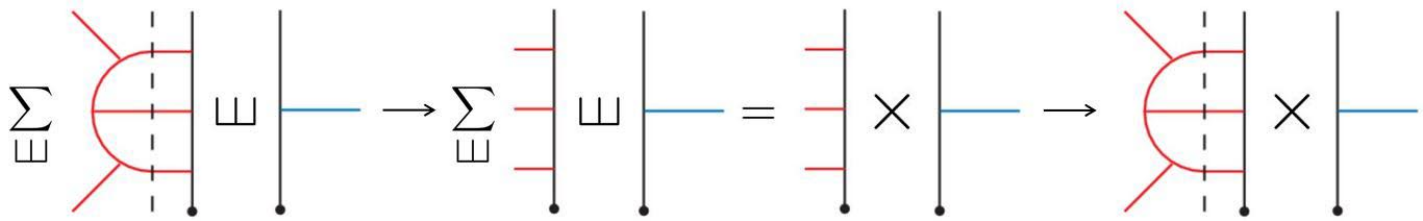} \\
  \caption{A $2$-loop example where the loops are entangled by sharing a propagator.}\label{Fig13}
\end{figure}

\begin{figure}
  \centering
   \includegraphics[width=14cm]{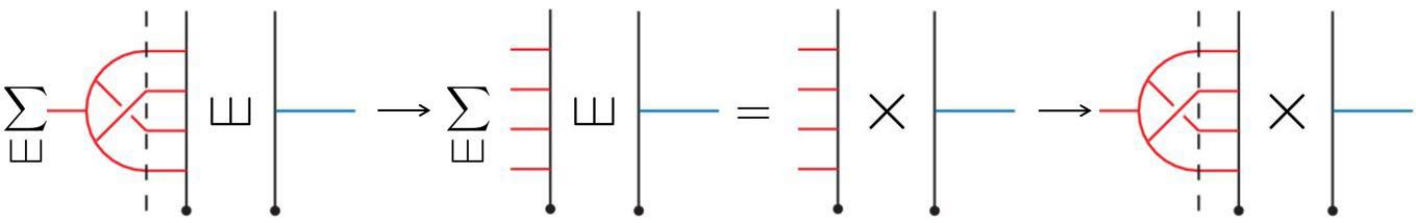} \\
  \caption{A $3$-loop non-planar example.}\label{Fig15}
\end{figure}

Fortunately, by relying on the method presented in section \ref{subsec-1loop-example}, the generalization of the SFASL to the case of entangled loops is equally straightforward. Take Fig.\ref{Fig13} as an example, in which two loops are entangled through the sharing of a single propagator. Following the same procedure as in the $1$-loop case, we first excise the $A$-side components, then apply the SFASL to the remainder---which is entirely equivalent to a tree-level diagram---and finally reattach the excised components. This yields the $2$-loop-level factorization behavior depicted in Fig.\ref{Fig13}. Evidently, the above argument applies equally to multi-loop configurations, even non-planar, such as that illustrated in Fig.\ref{Fig15}.
The factorization structure illustrated in the two examples, Figs.\ref{Fig13} and Fig.\ref{Fig15}, is guaranteed jointly by the kinematic condition $\ell\cdot k_b=0$ and by the convention of choosing the loop momentum from the $A$-side. It is easy to see that, based on these two constraints, the above generalization holds for an arbitrary number of loops. We emphasize again that, since the parts being cut off or spliced back are completely independent of the momenta from $\pmb B$, the cutting and splicing procedure does not have any substantive effect on the expressions corresponding to the diagrams.

\begin{figure}
  \centering
   \includegraphics[width=14cm]{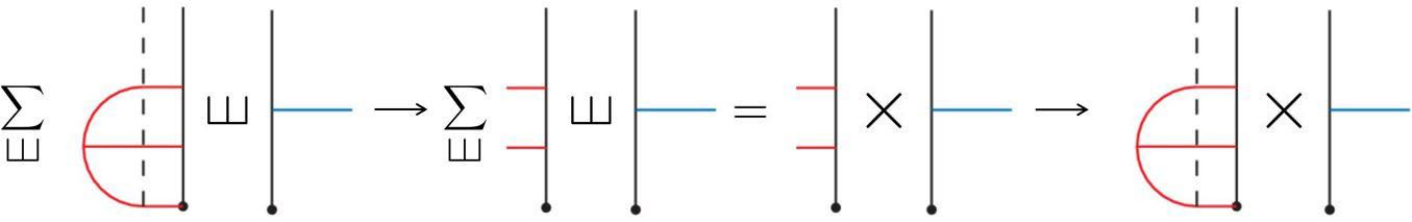} \\
  \caption{A $2$-loop example of scaleless terms.}\label{Fig14}
\end{figure}

Naturally, at higher loops, one also encounters the situation depicted in Fig.\ref{Fig11} and Fig.\ref{scaleless}: certain diagrams corresponding to scaleless integrals are removed, resulting in an incomplete set of shuffle permutations. The treatment of this case is the same with that at $1$-loop: one simply re-chooses the endpoint of $L_{(i,\bullet)}$ and then applies \eref{fac-propa-gen} to the new $L_{(i,\bullet)}$. The final result is a scaleless term with no physical contribution, which can be discarded directly. Fig.\ref{Fig14} provides a $2$-loop example, which is a direct generalization of Fig.\ref{Fig11} and Fig.\ref{scaleless}. It can be seen that the resulting factorization pattern in Fig.\ref{Fig14} contains a typical $2$-loop diagram corresponding to a scaleless integrand (the $A$-side part in the factorization). This is again a universal phenomenon due to the same reason as in the $1$-loop case. Whenever some diagrams in the full set of shuffle permutations contain a scaleless component, the left part of the shuffle permutation inherits that component. Hence, it ends up on the l.h.s. of the final factorization formula.

%

%%%%%%%%%%%%%%%%%%%%%%%%%%%%%%%
\subsection{Hidden zero and $2$-split of multi-loop Feynman integrand}
\label{subsec-higherloop-zeroandsplit}
%%%%%%%%%%%%%%%%%%%%%%%%%%%%%%%

By means of the SFASL for the multi-loop case discussed in the previous subsection, we can repeat all the discussions in section \ref{subsec-1loop-zeroandsplit} and generalize the hidden zeros and 2-split to an arbitrary number of loops.

First, the SFASL dictates that, under the kinematic condition \eref{kine-condi-Lloop-0}, the summation over shuffle permutations for a given division and a given $A$-side configuration yields a factorization pattern for Feynman diagrams that is completely analogous to that in Fig.\ref{Fig5}. At higher loops, the diagrams on both sides of $\times$ inside the curly brackets in Fig.\ref{Fig5} can contain highly complicated loop structures. Nevertheless, the factor outside the curly brackets remains $k_j^2$, which vanishes due to the on-shell condition. This completes the proof of the hidden zeros at multi-loop-level:
\bea
{\cal I}^{L-{\rm loop}}_n(1,\cdots,n)\,&\xrightarrow[]{\eref{kine-condi-Lloop-0}}&\,0\,,~~\label{zero-Lloop}
\eea
As at $1$-loop, the hidden zeros of multi-loop Feynman integrand are based on discarding a series of scaleless terms.

Next, again by means of SFASL, we can also show that, for a given division and a given $A$-side configuration, the factorization structure described in Fig.\ref{Fig6} and \eref{meaning-fig6} can be extended to higher loops.
The first premise of \eref{meaning-fig6}---two factorizations in \eref{two shuffles}---is ensured by the SFASL. Now we check the second premise, the factorization in \eref{key-fR}.
As in the $1$-loop case, when the loop lies on $L_{(i,j)}$, we need to multiply $f(R)$ and $f^A(R)$ in \eref{fR} by a factor that is more complicated than \eref{loop-fator}. However, no matter how complicated this factor may be, it does not affect the factorization structure expressed in \eref{key-fR}, since it does not receive any contribution from $\pmb B$. When the loop does not lie on $L_{(i,j)}$, we need to replace one or more BG currents in \eref{fR} by loop-level counterparts. However, this procedure likewise does not affect the validity of \eref{key-fR}. Since the factorization behaviors of both \eref{two shuffles} and \eref{key-fR} hold at higher loops, this guarantees that \eref{meaning-fig6} holds at higher loops. Summing over all divisions of $\pmb A$ and $\pmb B$, and all $A$-side configurations, $A$-replacements and $B$-replacements, we obtain
\bea
&&{\cal I}^{L-{\rm loop}}_n(1,\cdots,n)\,\xrightarrow[]{\eref{kine-condi-1loop-split}}\,\sum_{{\rm div}\pmb A}\,\sum_{{\rm div}\pmb B}\,\sum_{N_{(i,j)}}\sum_{N_A}\sum_{N_B}\,\delta(N_{(i,j)}+N_A+N_B-L)\,\sum_{\rm AC}\sum_{\rm AR}\sum_{\rm BR}\nn
&&~~~~~~~~~~~~~~~~~~~~~~~~~~~~~~~~\Big(\sum_{\shuffle(p,q)}\,\prod_{t=1}^{p+q}\,{1\over D_t^{(i,v)}}\Big)\,\Big(\sum_{\shuffle(m,l)}\,\prod_{t=1}^{m+l}\,{1\over D_t^{(j,v)}}\Big)\,
f(R)\nn
&=&\sum_{L_1=0}^L\,\Big[\sum_{{\rm div}\pmb A}\,\sum_{N_{(i,j)}}\sum_{N_A}\,\delta(N_{(i,j)}+N_A-L_1)\,\sum_{\rm AC}\sum_{\rm AR}\,\Big(\prod_{\a=1}^p\,{1\over s_{iA_1\cdots A_{\a}}}\Big)\,\Big(\prod_{\a=1}^m\,{1\over s_{iA'_1\cdots A'_{\a}}}\Big)\,f^A(R)\,\Big]\nn
&&~~~~~~~\times\,\Big[\sum_{{\rm div}\pmb B}\,\sum_{N_B}\,\delta(N_B-L+L_1)\,\sum_{BR}\,\Big(\prod_{\b=1}^q\,{1\over s_{iB_1\cdots B_{\b}}}\Big)\,\Big(\prod_{\b=1}^l\,{1\over s_{iB'_1\cdots B'_{\b}}}\Big)\,f^B(R)\,\Big]\,,~~~~\label{2split-Lloop-detail}
\eea
where the notations $\sum_{\rm AC}$, $\sum_{\rm AR}$, $\sum_{\rm BR}$, $N_{(i,j)}$, $N_A$, $N_B$ are explained below \eref{2split-1loop-detail}.
This serves as
the $2$-split of multi-loop Feynman integrands, which can be recast as
\bea
{\cal I}^{L-{\rm loop}}_n(1,\cdots,n)\,&\xrightarrow[]{\eref{kine-condi-Lloop-split}}&\,\sum_{L_1=0}^L\,\W{\cal I}^{L_1-{\rm loop}}_{n_1}(i,\pmb A,j,\kappa)\,\times\,\W{\cal I}^{(L-L_1)-{\rm loop}}_{n+3-n_1}(j,\pmb B(\kappa'),i)\,,~~\label{2split-Lloop}
\eea
where
\bea
&&\W{\cal I}^{L_1-{\rm loop}}_{n_1}(i,\pmb A,j,\kappa)\nn
&=&\sum_{{\rm div}\pmb A}\,\sum_{N_{(i,j)}}\sum_{N_A}\,\delta(N_{(i,j)}+N_A-L_1)\,\sum_{\rm AC}\sum_{\rm AR}\,\Big(\prod_{\a=1}^p\,{1\over s_{iA_1\cdots A_{\a}}}\Big)\,\Big(\prod_{\a=1}^m\,{1\over s_{iA'_1\cdots A'_{\a}}}\Big)\,f^A(R)\,,\nn
&&\W{\cal I}^{(L-L_1)-{\rm loop}}_{n+3-n_1}(j,\pmb B(\kappa'),i)\nn
&=&\sum_{{\rm div}\pmb B}\,\sum_{N_B}\,\delta(N_B-L+L_1)\,\sum_{BR}\,\Big(\prod_{\b=1}^q\,{1\over s_{iB_1\cdots B_{\b}}}\Big)\,\Big(\prod_{\b=1}^l\,{1\over s_{iB'_1\cdots B'_{\b}}}\Big)\,f^B(R)\,.
\eea

The $2$-split formula \eref{2split-Lloop} is the natural generalization of its $1$-loop counterpart \eref{2split-1loop}. In \eref{2split-Lloop}, each $\W{\cal I}^{L_1-{\rm loop}}_{n_1}$ is an $L_1$-loop object, and degenerates to the tree-level current when $L_1=0$.
Analogously, each $\W{\cal I}^{(L-L_1)-{\rm loop}}_{n+3-n_1}$ is an $(L-L_1)$-loop object, and reduces to the tree-level current when $L_1=L$. As in the $1$-loop case, each $\W{\cal I}^{L_1-{\rm loop}}_{n_1}$ carries an off-shell external leg $\kappa$ and lacks contributions from diagrams in which one or more loops lie on $L_{(\kappa,v)}$. Each $\W{\cal I}^{(L-L_1)-{\rm loop}}_{n+3-n_1}$ carries an off-shell external leg $\kappa'$, and omits diagrams those one or more loops lie on $L_{(i,j)}$, as well as diagrams that contain a $\kappa'$-bubble. Furthermore, each $\W{\cal I}^{(L-L_1)-{\rm loop}}_{n+3-n_1}$ contains no $\ell_m\cdot k_{B^{\rm sub}}$ in any denominator, for $\forall\,m\in\{1,\cdots,L\}$.

%%%%%%%%%%%%%%%%%%%%%%%%%%%%%%%
\section{Conclusion and discussion}
\label{sec-conclu}
%%%%%%%%%%%%%%%%%%%%%%%%%%%%%%%

In this paper, using a Feynman-diagram-based approach, we extend the hidden zeros and $2$-split of tree-level ${\rm Tr}(\phi^3)$ amplitudes to loop-level Feynman integrands, up to some scaleless integrals that have no physical contributions.
The loop-level hidden zeros and $2$-split obtained in our study are distinct from their counterparts in \cite{Backus:2025hpn,Arkani-Hamed:2024fyd}.
As can be seen, in our result the kinematic conditions for loop-level hidden zeros and $2$-split are very simple: it suffices to add $\ell\cdot k_{\hat{b}}$ to the tree-level kinematic conditions. Remarkably, the procedure for obtaining the $2$-split kinematic condition from the zero kinematic condition is exactly the same as that at tree-level. Compared with the $2$-split formula at tree-level, the $2$-split formula at loop-level represents a generalization: the $L$-loop integrand is expressed as a sum over $L+1$ terms, each of which exhibits a $2$-split structure. The physical interpretation of those terms in the $2$-split product that contain loops remains unclear at present, and we expect to address this issue in future work.

At tree level, a range of physical models exhibit hidden zeros and $2$-split structures in their amplitudes, including NLSM, YM, GR, etc., not limited to just ${\rm Tr}(\phi^3)$. A natural question, therefore, is whether the loop-level hidden zeros and $2$-split found in this paper can be generalized to other models. It is worth noting that such a generalization may be achievable using the Feynman-diagram-based approach developed in this paper. This is because our recent study shows that the tree-level diagram-based method described in section \ref{sec-tree} can be used to interpret the hidden zeros and $2$-split of tree-level NLSM, YM, and GR amplitudes, with only a slight generalization of the pattern of shuffle permutations (we will report this result in a separate paper). In light of this, it is reasonable to expect that this Feynman-diagram-based method can also be extended to other models at loop-level.

The method used in this paper is completely local, focusing only on the interaction vertices on a specific line. On the other hand, hidden zeros and $2$-split were originally discovered through surfaceology and the CHY formalism, both of which are global frameworks where locality is not manifest. An interesting question, therefore, is whether the loop-level hidden zeros and $2$-split found via the local method in this paper can be understood from these global frameworks. In particular, how can one establish a connection between local properties and global properties? This is another future direction worth pursuing.

%%%%%%%%%%%%%%%%%%%%%%%%%%%%%%%%%%%%%%%%%%%%%%
\section*{Acknowledgments}
%%%%%%%%%%%%%%%%%%%%%%%%%%%%%%%%%%%%%%%%%%%%%%%%%

This work is supported by NSFC under Grant No. 11805163.

%%%%%%%%%%%%%%%%%%%%%%%%%%%%%%%%%%%%%%%%%%%%%%%%%%%%

\bibliographystyle{JHEP}

\bibliography{reference}

\end{document}